\documentclass[fleqn,usenatbib]{mnras}

\usepackage[T1]{fontenc}

\DeclareRobustCommand{\VAN}[3]{#2}
\let\VANthebibliography\thebibliography
\def\thebibliography{\DeclareRobustCommand{\VAN}[3]{##3}\VANthebibliography}

\usepackage{graphicx}	% Including figure files
\usepackage{amsmath}	% Advanced maths commands
\usepackage{amssymb}	% Extra maths symbols
\usepackage{multicol}        % Multi-column entries in tables
\usepackage{bm}		% Bold maths symbols, including upright Greek
\usepackage{pdflscape}	% Landscape pages
\usepackage{makecell}
\usepackage{multirow}
\usepackage[normalem]{ulem}
\usepackage{orcidlink}
\usepackage{subcaption}
\usepackage{caption}

\usepackage{newtxtext,newtxmath}
\usepackage{xcolor}
\newcommand{\hmpc}{$h^{-1}$Mpc}
\newcommand{\ahod}{\textsc{AbacusHOD}}
\newcommand{\xirppi}{$\xi(r_\mathrm{p}, r_\pi)$\ }

\title[LRG HOD]{The DESI One-Percent Survey: Exploring the Halo Occupation Distribution of Luminous Red Galaxies and Quasi-Stellar Objects with \textsc{AbacusSummit} }

\author[Yuan et al.]{\parbox[t]{0.9\textwidth}{\vspace{-0.7cm}
Sihan Yuan,$^{1,2}$\thanks{E-mail: sihany@stanford.edu}\orcidlink{0000-0002-5992-7586}
Hanyu Zhang,$^{3}$\thanks{E-mail: hanyuz@phys.ksu.edu}\orcidlink{0000-0001-6847-5254}
Ashley J. Ross,$^{4}$
Jamie Donald-McCann,$^{5}$
Boryana Hadzhiyska,$^{6,7}$\orcidlink{0000-0002-2312-3121}
Risa H. Wechsler,$^{1,2}$\orcidlink{0000-0003-2229-011X}
Zheng Zheng,$^{8}$\orcidlink{0000-0003-1887-6732}
Shadab Alam,$^{9,10}$\orcidlink{0000-0002-3757-6359}
Violeta Gonzalez-Perez,$^{11,12}$\orcidlink{0000-0001-9938-2755}
Jessica Nicole Aguilar,$^{13}$
Steven Ahlen,$^{14}$\orcidlink{0000-0001-6098-7247}
Davide Bianchi,$^{15}$
David Brooks,$^{16}$
Axel de la Macorra,$^{17}$\orcidlink{0000-0002-1769-1640}
Kevin Fanning,$^4$\orcidlink{0000-0003-2371-3356}
Jaime E. Forero-Romero,$^{18}$\orcidlink{0000-0002-2890-3725}
Klaus Honscheid,$^{19,20}$
Mustapha Ishak,$^{21}$\orcidlink{0000-0002-6024-466X}
Robert Kehoe,$^{22}$
James Lasker,$^{22}$\orcidlink{0000-0003-2999-4873}
Martin Landriau,$^{13}$\orcidlink{0000-0003-1838-8528}
Marc Manera,$^{23}$\orcidlink{0000-0003-4962-8934}
Paul Martini,$^4$\orcidlink{0000-0002-4279-4182}
Aaron Meisner,$^{24}$\orcidlink{0000-0002-1125-7384}
Ramon Miquel,$^{23,25}$
John Moustakas,$^{26}$\orcidlink{0000-0002-2733-4559}
Seshadri Nadathur,$^5$
Jeffrey A. Newman,$^{27}$\orcidlink{0000-0001-8684-2222}
Jundan Nie,$^{28}$\orcidlink{0000-0001-6590-8122}
Will Percival,$^{29,30}$\orcidlink{0000-0002-0644-5727}
Claire Poppett,$^{13,31}$
Antoine Rocher,$^{32}$\orcidlink{0000-0003-4349-6424}
Graziano Rossi,$^{33}$
Eusebio Sanchez,$^{34}$\orcidlink{0000-0002-9646-8198}
Lado Samushia,$^{3}$
Michael Schubnell,$^{35}$
Hee-Jong Seo,$^{36}$\orcidlink{0000-0002-6588-3508}
Gregory Tarlé,$^{35}$\orcidlink{0000-0003-1704-0781}
Benjamin Alan Weaver,$^{24}$
Jiaxi Yu,$^{37}$
Zhimin Zhou,$^{28}$\orcidlink{0000-0002-4135-0977}
and Hu Zou$^{28}$\orcidlink{0000-0002-6684-3997}}
\vspace{0.3cm}
\\
\parbox{\textwidth}{
The authors' affiliations are shown in Appendix \ref{sec:affiliations}}.
\vspace{-0.5cm}}

\date{Accepted XXX. Received YYY; in original form ZZZ}

\pubyear{2015}

\begin{document}
\label{firstpage}
\pagerange{\pageref{firstpage}--\pageref{lastpage}}
\maketitle

\begin{abstract}
We present the first comprehensive Halo Occupation Distribution (HOD) analysis of the DESI One-Percent survey Luminous Red Galaxy (LRG) and Quasi Stellar Object (QSO) samples. We constrain the HOD of each sample and test possible HOD extensions by fitting the redshift-space galaxy 2-point correlation functions in $0.15<r<32\ h^{-1}$Mpc in a set of fiducial redshift bins. We use \textsc{AbacusSummit} cubic boxes at Planck 2018 cosmology as model templates and forward model galaxy clustering with the \textsc{AbacusHOD} package. We achieve good fits with a standard HOD model with velocity bias, and we find no evidence for galaxy assembly bias or satellite profile modulation at the current level of statistical uncertainty. For LRGs in $0.4 < z < 0.6$, we infer a satellite fraction of $f_\mathrm{sat} = 11\pm 1\%$, a mean halo mass of \smash{$\log_{10}\overline{M}_h=13.40^{+0.02}_{-0.02}$}, and a linear bias of \smash{$b_\mathrm{lin} = 1.93_{-0.04}^{+0.06}$}. 
For LRGs in $0.6 < z < 0.8$, we find $f_\mathrm{sat}=14\pm 1\%$, \smash{$\log_{10}\overline{M}_h=13.24^{+0.02}_{-0.02}$}, and \smash{$b_\mathrm{lin}=2.08_{-0.03}^{+0.03}$}. For QSOs, we infer \smash{$f_\mathrm{sat}=3^{+8}_{-2}\%$}, \smash{$\log_{10}\overline{M}_h= 12.65^{+0.09}_{-0.04}$}, and \smash{$b_\mathrm{lin} = 2.63_{-0.26}^{+0.37}$} in redshift range $0.8 < z < 2.1$. Using these fits, we generate a large suite of high fidelity galaxy mocks. We also study the redshift-evolution of the DESI LRG sample from $z = 0.4$ up to $z = 1.1$, revealing significant and interesting trends in mean halo mass, linear bias, and satellite fraction. 
%This paper is part of a series of detailed galaxy--halo connection studies on the DESI One-Percent Survey galaxy samples. 

\end{abstract}


%%%%%%%%%%%%%%%%%%%%%%%%%%%%%%%%%%%%%%%%%%%%%%%%%%

%%%%%%%%%%%%%%%%% BODY OF PAPER %%%%%%%%%%%%%%%%%%

\section{Introduction}
Galaxies are biased tracers of the underlying matter density field of the Universe, and their distribution is an important source of cosmological and astrophysical information. However, while the distribution of dark matter is readily modeled by gravitational collapse, the distribution of galaxies is significantly more complex due to nonlinear evolution and baryonic processes. Thus, to extract cosmology and galaxy physics from the observed galaxy distribution, it is critical to model the connection between galaxies and their underlying dark matter density field. 

A key piece of simplification in galaxy--dark matter connection modeling comes in what is known as the halo model, where simulations have shown that galaxies form and evolve in dense dark matter clumps known as halos \citep{1978White, 2002Cooray}. Within the halo model, we can empirically model the connection between galaxies and halos through a set of probabilistic models known as the Halo Occupation Distribution model \citep[HOD; e.g.][]{2000Peacock, 2001Scoccimarro, 2001White, 2002Berlind, 2003Berlind, 2005Zheng, 2007bZheng}. The HOD formalism has been highly successful in characterising magnitude-limited samples of bright galaxies in past galaxy redshift surveys \citep[e.g.][]{2011Zehavi, 2013Parejko, 2014Guo, 2015cGuo, 2016Rodriguez, 2020Alam, 2020Avila, 2021bYuan}. HOD studies are important not only because they reveal aspects of galaxy evolution physics and test assumptions of galaxy--dark matter connection \citep[e.g.][]{2019bLange, 2020Alam, 2021Yuan, 2022Wang, 2022Linke}, but also because they produce mocks that accurately reproduce the observed clustering and thus enable robustness tests of cosmology pipelines \citep[e.g.][]{2020Smith, 2021Rossi, 2021Alam}. Most recently, simulation-based forward modeling approaches have also utilised the flexibility of HODs to constrain cosmology from highly nonlinear scales that are otherwise inaccessible with standard analytical approaches \citep[e.g.][]{2021Lange, 2021Kobayashi, 2021Chapman, 2022bYuan, 2022Zhai}. 

The Dark Energy Spectroscopic Instrument (DESI) is a stage-IV spectroscopic galaxy survey with the primary goal of determining the nature of dark energy through the most precise measurement of the expansion history of the universe ever obtained \citep{2013Levi,2016DESI}. The baseline survey will obtain spectroscopic measurements of 40 million galaxies and quasars in a 14,000 deg$^2$ footprint in five years. This represents an order-of-magnitude improvement both in the volume surveyed and the number of galaxies measured over previous surveys. The DESI large-scale structure samples are divided into 4 target classes: the bright galaxy sample (BGS), the luminous red galaxies (LRG), the emission line galaxies (ELG), and the quasi-stellar objects (QSO). The auto- and cross-correlations of and between the four tracers probe the large-scale structure in increasing high redshift domains and combine to produce the most precise large-scale structure measurement from redshift $z = 0.1$ all the way to $z = 2.1$. Additionally, quasars that have redshifts greater than 2.1 are used as sight-lines for Lyman-$\alpha$ forest absorption, and the combination of ly$\alpha$-ly$\alpha$, ly$\alpha$-QSO, and QSO-QSO correlations probe large-scale structure to $z < 3.5$. 

The Early Data Release (EDR) of the DESI survey consists of data in the so-called One-Percent Survey, collected during the Survey Validation campaign~\citep[SV;][]{sv} before the start of the main survey operations. The One-Percent Survey covered 20 fields totalling 140 deg$^2$ with final target selection algorithms similar to those of the main survey \citep{2020Zhou, 2022Zhou, 2020Raichoor, 2022Raichoor, 2020Yeche, 2022Chaussidon, 2020Ruiz, 2022Hahn}. The One-Percent Survey reaches higher completeness than the main survey and produces the first clustering measurements from DESI. Specifically, more than 95$\%$ targets received fibers in the ELG sample, while more than 99$\%$ of targets in each of the BGS,
LRG, and QSO samples received fibers.

In this paper, we present a comprehensive HOD analysis of the DESI One-Percent Survey LRG and QSO samples.
% Equivalent analyses of the ELG and BGS samples are presented in \cite{Rocher} and \cite{Grove}, respectively. \sandy{potentially quote the uchuu papers too depending on how we phrase this}.
This paper is amongst a series of papers analyzing galaxy--halo connection models with DESI One-Percent Survey data. This paper addresses the more well understood samples of LRG and QSO, while the more novel ELG sample is analyzed in a dedicated paper \citep{2023Rocher}. In parallel, there are also several Subhalo-abundance matching (SHAM) analyses. Specifically, \cite{2023Prada} provides an overview of the \textsc{Uchuu}-based SHAM analyses \citep{2021Ishiyama}. 
%The results for individual tracers are presented in \textcolor{blue}{Ereza et al. in prep} for LRGs, \textcolor{blue}{Lasker et al. in prep} for ELGs and \textcolor{blue}{Rajeev et al. in prep} for QSOs and \textcolor{blue}{Smith et al. in prep} for BGS. 
\cite{2023Yu} presents SHAM analyses based on the \textsc{UNIT} simulation \citep{2019Chuang}. Beyond the single-tracer analyses, \cite{2023Gao} and \textcolor{blue}{Yuan et al. in prep} analyze the cross-correlation functions between the ELG and LRG tracers with multi-tracer SHAM and HOD models, respectively. These papers together present a significant variety of methodologies and mock products appropriate for a large scope of applications. 

This paper is structured as the following. In section~\ref{sec:data}, we introduce the observed samples and present their clustering measurements. In section~\ref{sec:sim} and \ref{sec:hod}, we introduce the simulation suite and the HOD models. In section~\ref{sec:results}, we present LRG fits on both the projected clustering measurements and the full-shape redshift-space clustering measurements, and present the corresponding model constraints. We also present a first analysis of the redshift evolution of the DESI LRG sample and the physical implications. We present the QSO fits in section~\ref{sec:qso}. In section~\ref{sec:mocks}, we present a series of mock products as a result of this analysis. Finally, we conclude in section~\ref{sec:conclude}.

Throughout this paper, we adopt the Planck 2018 $\Lambda$CDM cosmology, specifically the mean estimates of the Planck TT,TE,EE+lowE+lensing likelihood chains: $\Omega_c h^2 = 0.1200$, $\Omega_b h^2 = 0.02237$, $\sigma_8 = 0.811355$, $n_s = 0.9649$, $h = 0.6736$, $w_0 = -1$, and $w_a = 0$ \citep{2020Planck}.

\section{Data}
In this section, we describe the LRG and QSO samples and present their respective clustering measurements.  

DESI observed its One-Percent Survey as the third and final phase of its Survey Validation program in April and May of 2021. Observation fields were chosen to be in twenty non-overlapping `rosettes', where a high completeness was obtained by observing in each rosette at least 12 times. See \cite{sv} and \cite{edr} for more details.

Prior to beginning SV, 
the DESI instrument \citep{2022DESIinstrument} had proven its ability to simultaneously measure spectra at 5000 specific sky locations, with fibers placed accurately using robotic positioners populating the DESI focal plane \citep{DESIfocalplane}. During SV, the DESI data and operations teams' \citep{DESIsops} proved their ability to efficiently process the spectra through the DESI spectroscopic pipeline \citep{DESIspec}. Thus, DESI was able to start from an initial target list \citep{2023DESItarget} quickly obtain a highly complete One-Percent Survey.

The redshift measurements we use are available in the DESI EDR \citep{edr} 
\footnote{\url{https://data.desi.lbl.gov/public/edr/spectro/redux/fuji}}. These were input to the large-scale structure (LSS) catalogues, also described in the EDR \citep{edr}. Briefly, these LSS catalogues apply quality cuts to the data samples and provide matched random catalogues that trace the angular footprint and $dN/dz$ of the data, at a total density of 4.5$\times$10$^{4}\ $deg$^{-2}$. (\textcolor{blue}{Lasker et al. in prep}) describes how we simulated 128 alternative realisations of the DESI One-Percent Survey fiber assignment in order to encode via bits the realisations where each target was assigned and thus any joint probabilities of observation for a given set of targets. We use this information to determine the pairwise-inverse-probability \citep{2020Bianchi} weights to use in our clustering measurements. We further apply angular up-weighting (PIP+ANG) \citep{2020Bianchi}.  \cite{2020Mohammad} showed that this weighting scheme provides an unbiased clustering down to $0.1\ $\hmpc.

The One-Percent Survey LSS catalogues also include the so-called `FKP' \citep{FKP} weights in order to properly weight each volume element with respect to how each sample's number density changes with redshift,
\begin{equation}
    w_{\rm FKP} = 1/(1+n(z) P_0)
\end{equation}
where $n(z)$ is the weighted number per volume, and $P_0$ is a fiducial power-spectrum amplitude. We use $P_0=10^4\ (h^{-1}{\rm Mpc})^3$ for LRG and $P_0=6\times 10^3\ (h^{-1}{\rm Mpc})^3$ for QSO. For a detailed description of the weights and systematics treatment, we refer the readers to \cite{edr}.

\label{sec:data}
\subsection{DESI One-Percent Survey LRG and QSO samples}

% The targets for DESI are split into four target classes. In order of increasing average redshift these are: the Bright Galaxy Survey (BGS) targets, Luminous Red Galaxies (LRGs), Emission Line Galaxies (ELGs), and Quasars (QSOs). This paper will only analyze the LRG and QSO sample. The ELG sample will be analyzed in a separate paper (Rocher et al. in prep.), whereas the BGS sample will be analyzed in Grove. et al. in prep. 

% The LRG and QSO data sample studied in this paper was collected during the One-Percent survey of DESI. The latter was conducted at the end of the Survey Validation campaign 
% \sandy{cite sv paper}
% before the start of the main survey operations. The One-Percent survey covered 140 deg$^2$ with final target selection algorithms and depths similar to those of the main survey. 

The LRGs are an important type of galaxies for large-scale structure studies, and are specifically selected for observations due to two main advantages: 1) they are bright galaxies with the prominent 4000\AA\ break in their spectra, thus allowing for relatively easy target selection and redshift measurements; and 2) they are highly biased tracers of the large-scale structure, thus yielding a higher S/N per-object for the BAO measurement compared to typical galaxies. The LRG SV target selection is defined in \citet{2020Zhou}. The sample has a target density of 605 deg$^{-2}$ in $0.4 < z < 0.8$, significantly higher than previous LRG surveys \citep[BOSS and eBOSS][]{2013Dawson, 2016Dawson}, while the sample also extends to $z \sim 1$. Within EDR, the LRG main sample consists of 89,059 galaxies, 43,269 in the northern footprint and 45,790 in the southern footprint.

Quasi-stellar objects (a.k.a. Quasars, or QSOs) are the tracers of choice to study large-scale structures at high redshift due to the fact that they are some of the most luminous extragalactic sources. DESI aims to obtain spectra of nearly three million quasars, reaching limiting magnitudes $r \sim 23$ and an average density of $\sim$310 targets per deg$^2$. Within EDR, the QSO selection yields 24,182 quasars within redshift range $0.8 < z < 2.1$, and an additional 11,603 Ly-$\alpha$ quasars at higher redshift. For this study, we focus on the quasars at $z < 2.1$ which will be used for quasar clustering analysis.

\begin{figure}
    \hspace{-0.7cm}
    \includegraphics[width=0.54\textwidth]{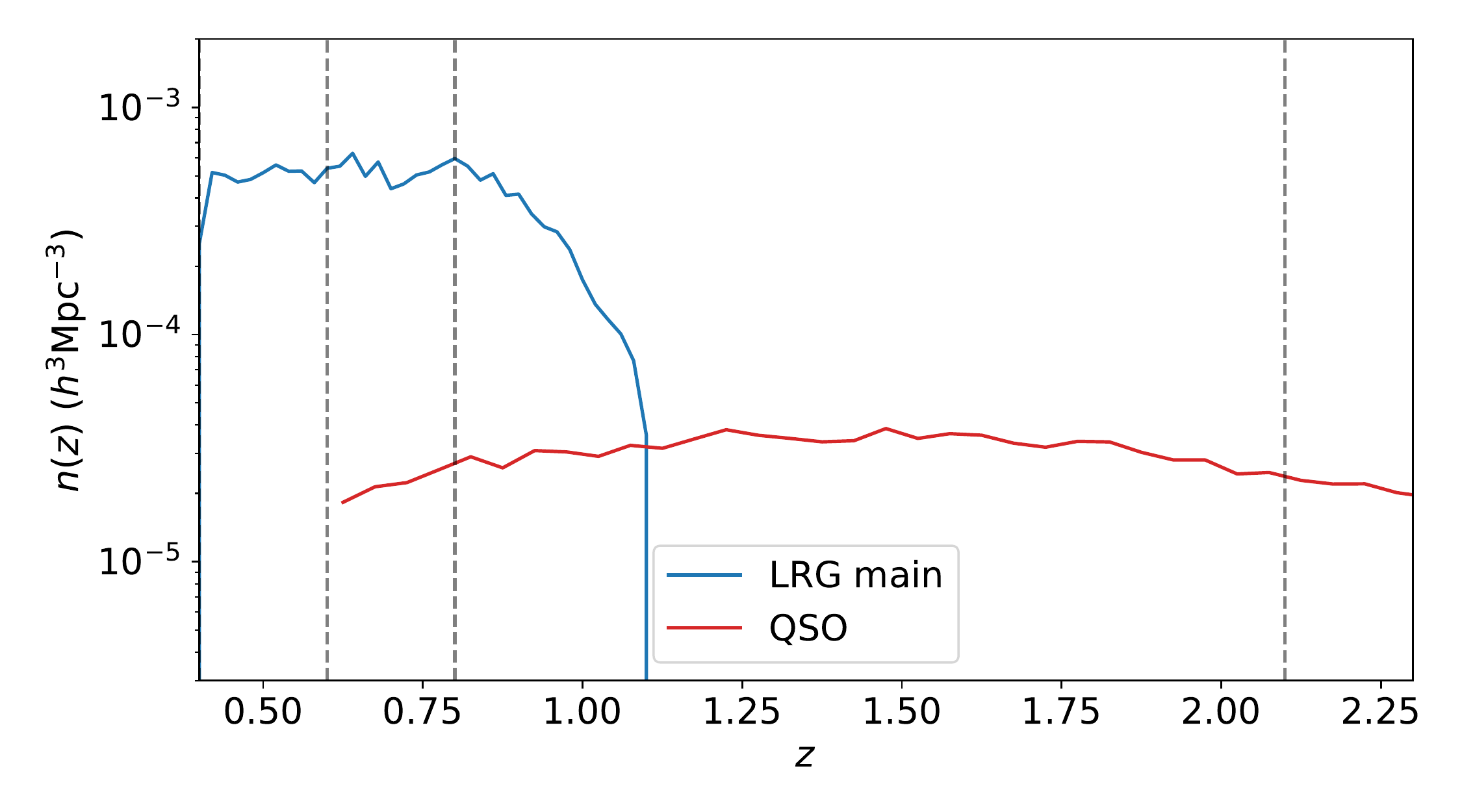}
    \vspace{-0.5cm}
    \caption{The DESI One-Percent Survey LRG and QSO mean number density as a function of redshift. The dashed vertical lines show the fiducial LRG redshift bin edges of $z = 0.6$, $z = 0.8$, and the maximum redshift we consider for the QSO sample $z = 2.1$.}
    \label{fig:nz}
\end{figure}

Figure~\ref{fig:nz} shows the DESI One-Percent Survey LRG and QSO mean density as a funtion of redshift $n(z)$. The vertical dashed lines correspond to fiducial bin edges defined for DESI cosmology studies. For the LRG sample, the number density remains fairly constant from $z = 0.4$ to $z = 0.8$ at approximately $5\times 10^{-4}\ h^{3}$Mpc$^{-3}$. At $z > 0.8$, the LRG density drops off quickly, suggesting increasing incompleteness and strong redshift evolution. For the fiducial HOD analysis presented in section~\ref{sec:results}, we examine the sample in two redshift bins: $0.4 < z < 0.6$ and $0.6 < z < 0.8$. Section~\ref{subsec:hodz} presents a preliminary analysis of the redshift evolution of the LRGs at $z > 0.8$.

The QSO sample delivers roughly constant number density from $z = 0.8$ to $z = 2.1$, at $2\times 10^{-5}\ h^{3}$Mpc$^{-3}$. In this analysis, we treat this entire redshift range as one single bin to achieve a reasonably large sample size for clustering measurements. We nevertheless expect at least some degree of redshift evolution, but we defer the analysis of QSO redshift evolution to a future paper when a larger sample becomes available.

\subsection{Clustering measurements}
For this analysis, we consider the 2-point correlation function (2PCF) as our summary statistic of the galaxy clustering. We start by introducing the redshift-space 2PCF $\xi(r_p, r_\pi)$, which can be computed using the \citet{1993Landy} estimator:
\begin{equation}
    \xi(r_p, r_\pi) = \frac{DD - 2DR + RR}{RR},
    \label{equ:xi_def}
\end{equation}
where $DD$, $DR$, and $RR$ are the normalised numbers of data-data, data-random, and random-random pair counts in each bin of $(r_p, r_\pi)$. $r_p$ and $r_\pi$ are transverse and line-of-sight (LoS) separations in comoving units. The redshift-space $\xi(r_p, r_\pi)$ in principle represents the full information content of the 2PCF. The dependence on transverse separation $r_p$ describes the transition from 1-halo clustering to 2-halo clustering, whereas the dependence on LoS separaton $r_\pi$ details the velocity distributions and the small-scale finger-of-god effect. \cite{2021bYuan} showed that the $\xi(r_p, r_\pi)$ on small scales yield strong constraints on the HOD. In this paper, we consider $\xi(r_p, r_\pi)$ as our primary summary data vector for constraining the LRG and QSO HOD. 

However, $\xi(r_p, r_\pi)$ is often compressed to the projected galaxy 2PCF $w_p$, which is the line-of-sight integral of $\xi(r_p, r_\pi)$,
\begin{equation}
w_p(r_p) = 2\int_0^{r_{\mathrm{\pi, max}}} \xi(r_p, r_\pi)dr_\pi,
\label{equ:wp_def}
\end{equation}
By definition, $w_p$ is strictly less informative than $\xi(r_p, r_\pi)$ as it loses out on the velocity information that is encoded in the LoS clustering. However, $w_p$ also offer several key advantages: it is easy to visualise as a 1D function; it is easy to obtain covariance matrix for; analyzing $w_p$ avoids the complexities of modeling galaxy velocities. For these reasons, we present $w_p$-only results alongside the $\xi(r_p, r_\pi)$ results in the following analysis. 
 
% \sandy{Talk about pip weights and corrections to 2pcf when that becomes available}
% The clustering catalogues of one-percent data provide the FKP weight \textit{WEIGHT\_FKP} that minimises the cosmic variance \citep{FKP1994}
% \begin{equation}
%     w_{\rm FKP} = \frac{1}{1+\overline{n}(z)P_0},
% \end{equation}
% where $\overline{n}(z)$ is the average number density at redshift $z$ and $P_0$ is the amplitude of the observed power spectrum at $k\approx {0.15}\, h\,{\rm Mpc}^{-1}$. $P_0={10000}\, h^{-3}\,{\rm Mpc}^{3}$, $4000\, h^{-3}\,{\rm Mpc}^{3}$, $6000\, h^{-3}\,{\rm Mpc}^{3}$ for LRGs, ELGs and QSOs respectively. 

% The pairwise-inverse-probability \citep{bianchi17} and angular up-weighting (PIP+ANG) \citep{percival17} weights can also be used to correct the effect of fibre collision. \citep{Mohammad_2020} prove that this weighting scheme provides an unbiased clustering down to 0.1$\hmpc$. So the observational clustering measurements of this study apply PIP+ANG weight. 

\begin{figure*}
    \hspace{-0.7cm}
    \begin{subfigure}[b]{0.45\textwidth}
         \centering
         \includegraphics[width=\textwidth]{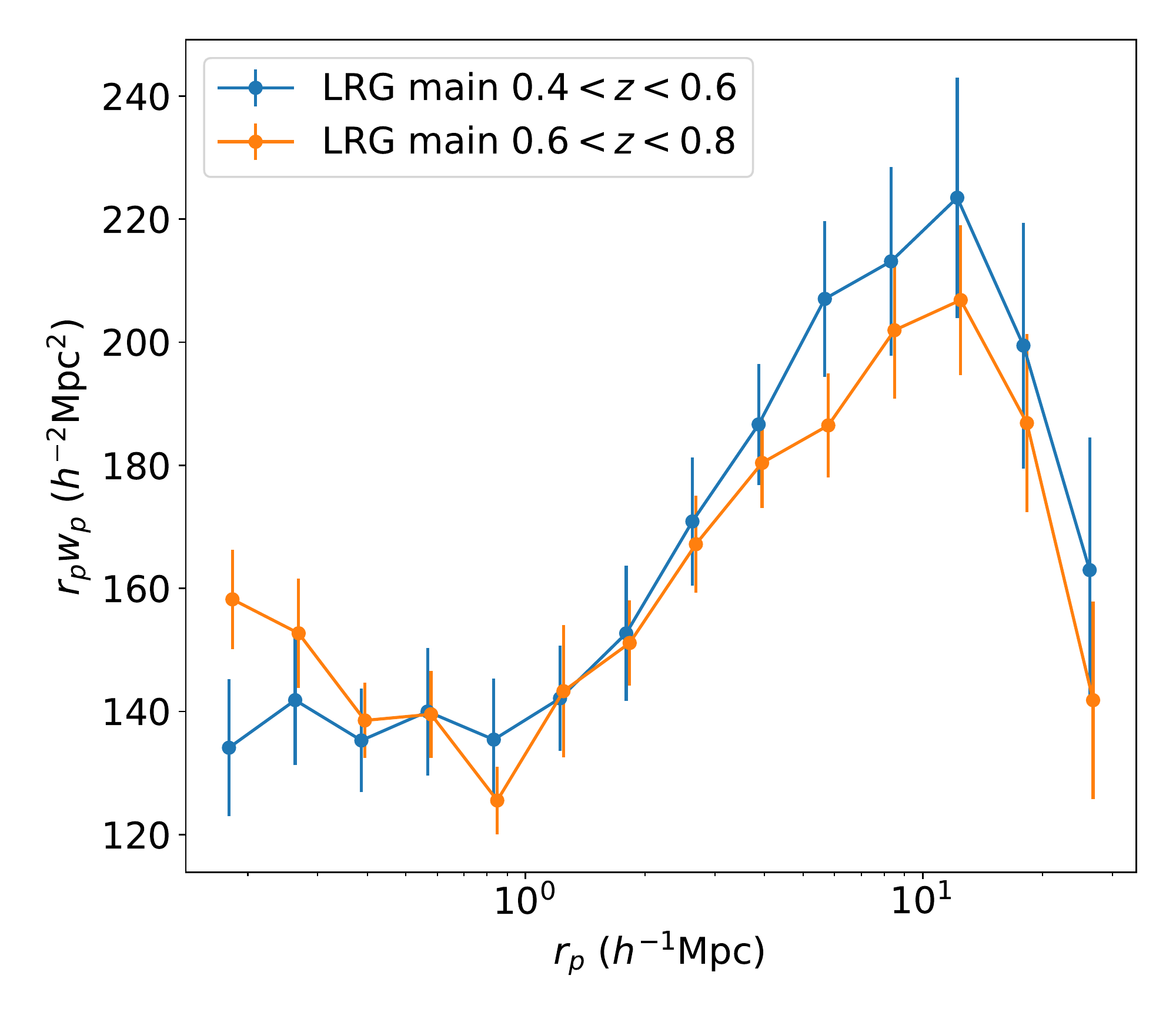}
        \vspace{-0.7cm}
         \caption{LRG sample}
         \label{fig:wp_lrg}
     \end{subfigure}
    \begin{subfigure}[b]{0.45\textwidth}
         \centering
         \includegraphics[width=\textwidth]{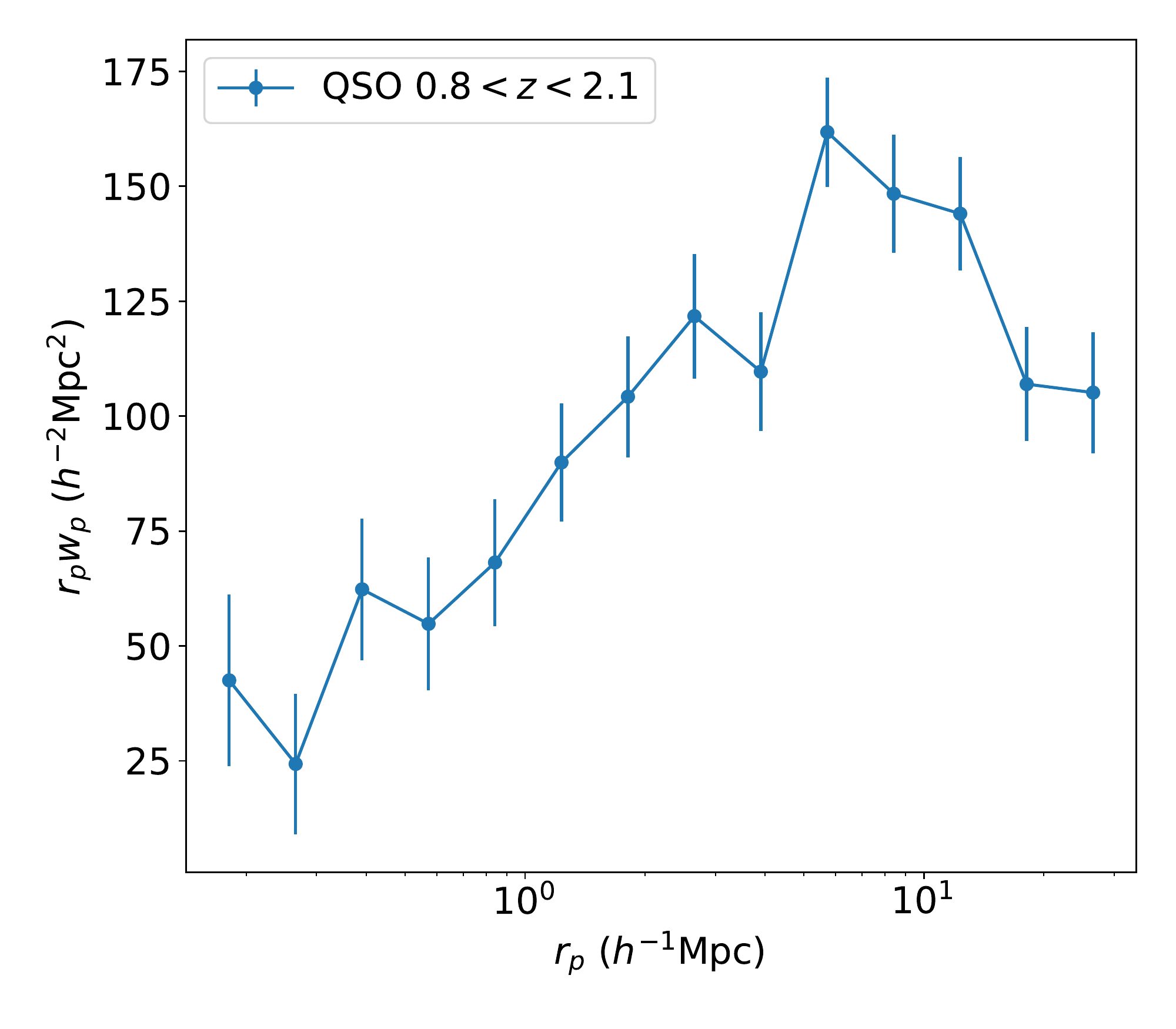}
        \vspace{-0.7cm}
         \caption{QSO sample}
         \label{fig:wp_qso}
     \end{subfigure}
    \caption{The DESI One-Percent Survey LRG and QSO projected auto-correlation functions. Here we are only showing LRG clustering in two fiducial redshift bins: $0.4 < z < 0.6$ and $0.6 < z < 0.8$. For QSOs, we consider one large redshift bin $0.8 < z < 2.1$ to achieve a reasonable sample size. }
    \label{fig:wp}
\end{figure*}

Figure~\ref{fig:wp} shows the projected auto-correlation function of the DESI One-Percent Survey LRG and QSO samples, using the fiducial redshift bins we defined above. Throughout the rest of this analysis, we adopt 14 logrithmic bins along the projected separation $r_p$ from $0.15\ h^{-1}$Mpc to $32\ h^{-1}$Mpc. The projected scale range is designed to capture both the 1-halo regime and the 1 to 2 halo transition regime, while limiting our exposure to large scale modes due to the small footprint in the One-Percent Survey. Along the Line-of-sight (LoS) direction, we adopt a linear binning scheme from 0 to $32\ h^{-1}$Mpc with bin size $\Delta r_\pi = 4\ h^{-1}$Mpc to capture the structure of the finger-of-god effect without blowing up the size of the data vector. For $w_p$, we set $r_\mathrm{\pi, max} = 32\ h^{-1}$Mpc. The redshift multipole measurements are visualised in later figures (Figure~\ref{fig:multipoles} and \ref{fig:multipoles_qso}). 
The errorbars displayed alongside the data measurements are calculated with 128 jackknife regions of the One-Percent Survey footprint. All clustering measurements on DESI One-Percent Survey data are done using the
\textsc{pycorr} package \footnote{\url{https://github.com/cosmodesi/pycorr}} \citep{jackknife2022}.

\section{Simulations}
\label{sec:sim}

To model the underlying dark matter density field, we use the \textsc{AbacusSummit} simulation suite, which is a set of large, high-accuracy cosmological N-body simulations using the \textsc{Abacus} N-body code \citep{2021Maksimova, 2019Garrison, 2021bGarrison}. This suite is designed to meet the Cosmological Simulation Requirements of DESI. \textsc{AbacusSummit} consists of over 150 simulations, containing approximately 60 trillion particles at 97 different cosmologies. A base simulation box contains $6912^3$ particles within a $(2\ h^{-1}$Gpc$)^3$ volume, which yields a particle mass of $2.1 \times 10^9\ h^{-1}M_\odot$. \footnote{For more details, see \url{https://abacussummit.readthedocs.io/en/latest/abacussummit.html}}

The simulation output is organised into discrete redshift snapshots. Specifically, we use the $z=0.5$ and $z=0.8$ snapshots for LRGs at $z<0.8$, and the $z=0.8$ and $z=1.1$ snapshots for LRGs at $z>0.8$. % \sout{ For this analysis, we use the $z = 0.5$ and $z = 0.8$ snapshots for LRGs in redshift bin $0.4 < z < 0.6$ and $0.6 < z < 0.8$, respectively.}. 
For the QSO analysis, due to the very limited sample size, we choose not to divide the sample into multiple redshift bins. Instead, we use the $z = 1.4$ snapshot for the single redshift bin $0.8 < z < 2.1$. A more nuanced analysis of the QSO sample is planned when more data become available. All fits are done at Planck cosmology with the \verb+AbacusSummit_base_c000_ph000+ box. 

% The set of HOD fits presented in this paper are done primarily using the $z = 0.5$ and $z = 0.8$ slices of the \verb+AbacusSummit_base_c000_ph000+ box, which is $(2h^{-1}$Gpc$)^3$ in volume and adopts the Planck 2018 $\Lambda$CDM cosmology ($\Omega_c h^2 = 0.1200$, $\Omega_b h^2 = 0.02237$, $\sigma_8 = 0.811355$, $n_s = 0.9649$, $h = 0.6736$, $w_0 = -1$, and $w_a = 0$). 

The dark matter halos are identified with the {\sc CompaSO} halo finder, which is a highly efficient on-the-fly group finder specifically designed
for the \textsc{AbacusSummit} simulations \citep{2021Hadzhiyska}. 
{\sc CompaSO} builds on the existing 
spherical overdensity (SO) algorithm
by taking into consideration the tidal radius
around a smaller halo before competitively
assigning halo membership to the particles
in an effort to more effectively deblend halos.
Among other features, the {\sc CompaSO} finder also
allows for the formation of new halos on the 
outskirts of growing halos, which alleviates
a known issue of configuration-space halo 
finders of failing to identify halos close to
the centers of larger halos. 
We also run a post-processing ``cleaning'' procedure that leverages the halo merger trees to ``re-merge'' a subset of halos. This is done both to remove over-deblended halos in the spherical overdensity finder, and to intentionally merge physically associated halos that have merged and then physically separated \citep{2021Bose}. 
%An example of such dissociation is what is known as splashback \citep[e.g.][]{2014Diemer, 2015bMore, 2016More}, where halos that were once part of a larger halos have since exited following at least one orbital passage within their former hosts. In \citet{2021Bose}, we find that remerging such halos signicantly improves the fidelity of the halo catalogue, and the resulting ``cleaned'' halo catalogue achieves significantly better fits on data in an HOD analysis. The fits presented in later sections of this paper are carried out with the cleaned halo catalogues. 

% maybe also talk about lightcone products if needed 
In addition to periodic boxes, the simulation suite also provides a set of simulation lightcones at fiducial cosmology \citep[][]{2022Hadzhiyska}. The basic algorithm associates the halos from a set of coarsely-spaced snapshots with their positions at the time of light-cone crossing by matching halo particles to on-the-fly light cone particles. The resulting halo catalogues are reliable at $M_\mathrm{halo} > 2.1\times 10^{11}\ h^{-1}M_\odot$, more than sufficient for LRGs and QSOs. As part of the data products, we take the best-fit HODs across different redshift snapshots and construct redshift-dependent LRG mocks on the 25 base lightcones. Each lightcone covering an octant of the sky ($\sim 5156$ deg$^2$) up to $z\sim 0.8$. We clarify that in this analysis, we only use the cubic boxes to conduct our analysis, the lightcones are only used to produce redshift-dependent mocks as part of the data products. 

\section{Halo Occupation Distribution (HOD)}
\label{sec:hod}
To propagate the simulated matter density field to galaxy distributions, we adopt the Halo Occupation Distribution model (HOD), which probabilistically populate dark matter halos with galaxies according to a set of halo properties. Statistically, the HOD can be summarised as a probabilitistic distribution $P(n_g|\boldsymbol{X}_h)$, where $n_g$ is the number of galaxies of the given halo, and $\boldsymbol{X}_h$ is some set of halo properties.

In the vanilla HOD model, halo mass is assumed to be the only relevant halo property $\boldsymbol{X}_h = {M_h}$ \citep{2005Zheng, 2007bZheng}. This vanilla HOD separates the galaxies into central and satellite galaxies, and assumes the central galaxy occupation follows a Bernoulli distribution whereas the satellites follow a Poisson distribution. Beyond the vanilla model, galaxy occupation can also depend on secondary halo properties beyond halo mass, an phenomenon commonly referred to as galaxy assembly bias or galaxy secondary bias \citep[See][for a review]{2018Wechsler}. While galaxy assembly bias is well physically motivated, many studies have looked for it both in simulations and data \citep[e.g.][]{2002Wechsler, 2007Croton, 2007Gao, 2016Lin, 2020Hadzhiyska, 2020Xu, 2021Xu, 2021Delgado, 2022Salcedo, 2022Yuan, 2022Wang} with mixed results. 
%In this analysis, we test the inclusion of two different flavors of galaxy assembly bias.  

For this analysis, we use the \ahod\ code to find best-fit HODs and sample HOD posteriors. \ahod\ is a highly efficient HOD implementation that enables a large set of HOD extensions \citep[][]{2021bYuan}. The code is publicly available as a part of the \textsc{abacusutils} package at \url{https://github.com/abacusorg/abacusutils}. Example usage can be found at \url{https://abacusutils.readthedocs.io/en/latest/hod.html}. 

% \subsection{LRG}
\subsection{Baseline model}
For a LRG sample, the HOD is well approximated by a vanilla model given by (originally shown in \citealt{2007bZheng} and referred to as Zheng07 or vanilla later in the text):
\begin{align}
    \bar{n}_{\mathrm{cent}}^{\mathrm{LRG}}(M) & = \frac{f_\mathrm{ic}}{2}\mathrm{erfc} \left[\frac{\log_{10}(M_{\mathrm{cut}}/M)}{\sqrt{2}\sigma}\right], \label{equ:zheng_hod_cent}\\
    \bar{n}_{\mathrm{sat}}^{\mathrm{LRG}}(M) & = \left[\frac{M-\kappa M_{\mathrm{cut}}}{M_1}\right]^{\alpha}\bar{n}_{\mathrm{cent}}^{\mathrm{LRG}}(M),
    \label{equ:zheng_hod_sat}
\end{align}
where the five vanilla parameters characterising the model are $M_{\mathrm{cut}}, M_1, \sigma, \alpha, \kappa$. $M_{\mathrm{cut}}$ sets the minimum halo mass to host a central galaxy. $M_1$ roughly sets the typical halo mass that hosts one satellite galaxy. $\sigma$ controls the steepness of the transition from 0 to 1 in the number of central galaxies. $\alpha$ is the power law index on the number of satellite galaxies. $\kappa M_\mathrm{cut}$ gives the minimum halo mass to host a satellite galaxy.
We have added a modulation term $\bar{n}_{\mathrm{cent}}^{\mathrm{LRG}}(M)$ to the satellite occupation function to mostly remove satellites from halos without centrals\footnote{There is evidence that such central-less satellites may exist in a realistic stellar-mass selected catalogue \citep[][]{2019Jimenez}. We include this term for consistency with previous works, but it should have minimal impact on clustering.}. We have also included an incompleteness parameter $f_\mathrm{ic}$, which is a downsampling factor controlling the overall number density of the mock galaxies. This parameter is relevant when trying to match the observed mean density of the galaxies in addition to clustering measurements. By definition, $0 < f_\mathrm{ic}\leq 1$.

For QSOs, we adopt essentially the same HOD model except we remove the central modulation term in the satellite occupation as there is no evidence that the existence of satellite QSOs are strongly associated with central QSOs. Thus, for satellite QSOs, we have 
\begin{align}
    \bar{n}_{\mathrm{sat}}^{\mathrm{QSO}}(M) & = \left[\frac{M-\kappa M_{\mathrm{cut}}}{M_1}\right]^{\alpha}.
    \label{equ:zheng_hod_sat_qso}
\end{align}

In addition to determining the number of galaxies per halo, the standard HOD model also dictates the position and velocity of the galaxies. In the vanilla model, the position and velocity of the central galaxy are set to be the same as those of the halo center, specifically the L2 subhalo center-of-mass for the {\sc CompaSO} halos \citep[see ][]{2021Hadzhiyska}. For the satellite galaxies, they are randomly assigned to halo particles with uniform weights, each satellite inheriting the position and velocity of its host particle. 

Because we are modeling the full-shape \xirppi, we also include an additional level of flexibility in the velocity model known as velocity bias in the baseline model. Velocity bias essentially parametrises any biases the velocities of the central and satellite galaxies relative to their respectively host halos and particles. This is shown to to be a necessary ingredient in modeling BOSS LRG redshift-space clustering on small scales \citep[e.g.][]{2015aGuo, 2021bYuan}. Velocity bias has also been identified in hydrodynamical simulations and measured to be consistent with observational constraints \citep[e.g.][]{2022Yuan, 2017Ye}. 

We parametrise velocity bias through two additional parameters: 
\begin{itemize}
    \item \texttt{$\alpha_\mathrm{vel, c}$} is the central velocity bias parameter, which modulates the peculiar velocity of the central galaxy relative to the halo center along the LoS. Specifically in this model, the central galaxy velocity along the LoS is thus given by 
    \begin{equation}
        v_\mathrm{cent, z} = v_\mathrm{L2, z} + \alpha_\mathrm{vel, c} \delta v(\sigma_{\mathrm{LoS}}),
        \label{equ:alphac}
    \end{equation}
    where $v_\mathrm{L2, z}$ denotes the LoS component of the central subhalo velocity, $\delta v(\sigma_{\mathrm{LoS}})$ denotes the Gaussian scatter, and $\alpha_\mathrm{vel, c}$ is the central velocity bias parameter. By definition, $\alpha_\mathrm{vel, c} = 0$ corresponds to no central velocity bias. We also define $\alpha_\mathrm{vel, c}$ as non-negative, as negative and positive $\alpha_c$ are fully degenerate observationally. 
    \item \texttt{$\alpha_\mathrm{vel, s}$} is the satellite velocity bias parameter, which modulates how the satellite galaxy peculiar velocity deviates from that of the local dark matter particle. 
    Specifically, the satellite velocity is given by 
    \begin{equation}
        v_\mathrm{sat, z} = v_\mathrm{L2, z} + \alpha_\mathrm{vel, s} (v_\mathrm{p, z} - v_\mathrm{L2, z}),
        \label{equ:alpha_s}
    \end{equation}
    where $v_\mathrm{p, z}$ denotes the line-of-sight component of particle velocity, and $\alpha_\mathrm{vel, s}$ is the satellite velocity bias parameter.
    $\alpha_\mathrm{vel, s} = 1$ indicates no satellite velocity bias, i.e. satellites perfectly track the velocity of their underlying particles. 
    
\end{itemize}

To summarise, the baseline HOD model for both LRGs and QSOs is fully specified with the following 8 parameters: (1) 5 vanilla HOD parameters $M_{\mathrm{cut}}$, $M_1$, $\sigma$, $\alpha$, $\kappa$; (2) an incompleteness parameter $f_\mathrm{ic}$; (3) velocity bias parameters $\alpha_\mathrm{vel, c}$ and $\alpha_\mathrm{vel, s}$.

\subsection{Model extensions}

\ahod\ also enables additional physically motivated HOD extensions. In the following analysis, we will test whether the data favor the inclusion of such extensions. We summarise the relevant extensions for LRGs here and refer the readers to \cite{2021bYuan} for more details:
\begin{itemize}

    \item \texttt{$A_\mathrm{cent}$} or \texttt{$A_\mathrm{sat}$} are the concentration-based secondary bias parameters for centrals and satellites, respectively. Also known as galaxy assembly bias parameters. $A_\mathrm{cent} = 0$ and $A_\mathrm{sat} = 0$ indicate no concentration-based secondary bias in the centrals and satellites occupation, respectively. A positive $A$ indicates a preference for lower concentration halos, and vice versa. 
    \item \texttt{$B_\mathrm{cent}$} or \texttt{$B_\mathrm{sat}$} are the environment-based secondary bias parameters for centrals and satellites, respectively. The environment is defined as the mass density within a $r_\mathrm{env} = 5\ h^{-1}$Mpc tophat of the halo center, excluding the halo itself. $B_\mathrm{cent} = 0$ and $ B_\mathrm{sat} = 0$ indicate no environment-based secondary bias. A positive $B$ indicates a preference for halos in less dense environments, and vice versa. 
    \item \texttt{$s$} is the satellite profile bias parameter, which modulates how the radial distribution of satellite galaxies within haloes deviate from the radial profile of the halo (potentially due to baryonic effects). $s = 0$ indicates no radial bias, i.e. satellites are uniformly assigned to halo particles. $s > 0$ indicates a more extended (less concentrated) profile of satellites relative to the halo, and vice versa. 
\end{itemize}

For this paper, we will add each of these extensions on to the 8-parameter baseline HOD model and conduct fits on the data. We compare the fits to study whether any of these extensions are favored. However, we only test these extensions on the LRG sample. While similar extensions might also apply for QSOs, we lack the sufficient sample size to meaningfully constrain such effects. 

\subsection{Redshift-space distortion}

Having generated the mock galaxy catalogues with each HOD prescription, we need to compute the 2PCF to compare to the data. However, because the data is in redshift space, meaning the observed LoS positions of galaxies are shifted by their peculiar velocity divided by the Hubble constant, we need to incorporate this effect in our model too. Thus, we impose redshift-space distortion (RSD) on the $z$-axis positions of the mock galaxies by amount
\begin{equation}
    Z_{\rm redshift}=Z_{\rm real}+\frac{v_{\rm pec,Z}(1+z)}{H(z)}, 
    \label{real-redshift}
\end{equation} 
where $Z_{\rm real}$ and $Z_{\rm redshift}$ are the real and redshift-space $z$-axis positions of the galaxies. $v_{\rm pec,Z}$ is the galaxy peculiar velocity projected along the $z$-axis. $H(z)$ is the Hubble parameter at redshift $z$. The $1+z$ scaling converts the coordinates into comoving units. 

Finally, we compute the model predicted \xirppi directly from mocks, assuming $z$-axis as the LoS direction. We use the grid-based 2PCF calculator \textsc{Corrfunc} \citep{2020Sinha} for efficiency.

\section{Likelihood model and covariance matrix}
\label{sec:cov}

To perform the subsequent optimisations and sampling of the HOD parameters, we need to construct a likelihood function. In this analysis, we assume a simple Gaussian likelihood and utilise the $\chi^2$ statistic:
\begin{equation}
       \chi^2_{\xi}  = (\xi_{\mathrm{model}} - \xi_{\mathrm{data}})^T \boldsymbol{C}^{-1}(\xi_{\mathrm{model}} - \xi_{\mathrm{data}}),
       \label{equ:chi2xi}
\end{equation}
where the $\xi_{\mathrm{model}}$ is the model predicted $\xi(r_p, \pi)$ and $\xi_\mathrm{data}$ is the DESI measurement. $\boldsymbol{C}$ is the covariance matrix. 
% The $\xi_\mathrm{data}$ was described in section~\ref{sec:data} whereas $\xi_{\mathrm{model}}$ can be straightforwardly computed from the HOD mocks via any 2PCF calculator (e.g. \textsc{Corrfunc} \citet{2020Sinha}). We present the covariance matrix model later in this section. 

We also include in the likelihood model an additional term related to the mean number density of the sample, 
\begin{equation}
   \chi^2_{n_g} = \begin{cases}
   \left(\frac{n_{\mathrm{mock}} - n_{\mathrm{data}}}{\sigma_{n}}\right)^2 & (n_{\mathrm{mock}} < n_{\mathrm{data}}) \\
   0 & (n_{\mathrm{mock}} \geq n_{\mathrm{data}}).
   \end{cases}
   \label{equ:chi2ng}
\end{equation}
$\sigma_n$ is the uncertainty of the galaxy number density. The $\chi^2_{n_g}$ is a half normal around the observed number density $n_\mathrm{data}$. When the mock number density is less than the data number density $(n_{\mathrm{mock}} < n_{\mathrm{data}})$, we set the completeness to $f_\mathrm{ic} = 1$ and give a Gaussian-type penalty on the difference between $n_{\mathrm{mock}}$ and $n_{\mathrm{data}}$. When the mock number density is higher than data number density $(n_{\mathrm{mock}} \geq n_{\mathrm{data}})$, then we set $f_{\mathrm{ic}} = n_{\mathrm{data}}/n_{\mathrm{mock}}$ such that the mock galaxies catalogue is uniformly downsampled to match the data number density. In this case, we impose no penalty. This definition of $\chi^2_{n_g}$ allows for incompleteness in the observed galaxy sample while penalising HOD models that produce insufficient galaxy number density. 
For the rest of this paper, we assume $\sigma_n = 0.1n_{\mathrm{data}}$.

Finally, the full $\chi^2$ is given by
\begin{equation}
\chi^2  = \chi^2_{\xi} + \chi^2_{n_g}.
\label{equ:chi2tot}
\end{equation}

To obtain the covariance matrix, one could divide the observed sample into jackknife regions and compute the clustering in each assuming they are independent realisations. However, given the finite size of the One-Percent Survey footprint and the relatively large number of bins in the \xirppi statistic, the resulting jackknife covariances are noisy and close to singular. Instead, we opt to use the 1800 $500\ $\hmpc\ boxes with varying phases in the \textsc{AbacusSummit} suite to generate mock-based covariance matrices. 
Specifically, each small box shares the same particle resolution as the base boxes and is $500\ h^{-1}$Mpc per side, which is sufficient for the scales we analyze. 

First, we generate mocks on the 1800 boxes that produce the same clustering as that measured in data. Specifically, we take a baseline HOD model and fit the observed \xirppi with just the jackknife errors measured on the data, assuming all off-diagonal terms in the covariance matrices are zero. We achieve good fits for both tracers and in all redshift bins. We do not present the values of this fit to avoid confusion with the final ``full covariance'' fit presented in Table~\ref{tab:LRG_fit1}, but the parameter values are consistent with the ``full covariance'' fits. 
We then take the best-fit HOD and populate the 1800 covariance boxes, from which compute the covariance matrices. Finally, we renormalise the covariance matrix by keeping the mock-based correlation matrix and using the data-based jackknife diagonal errors to convert the correlation matrix to the final covariance matrix.

\begin{figure}
    \hspace{-0.4cm}
    \includegraphics[width=0.49\textwidth]{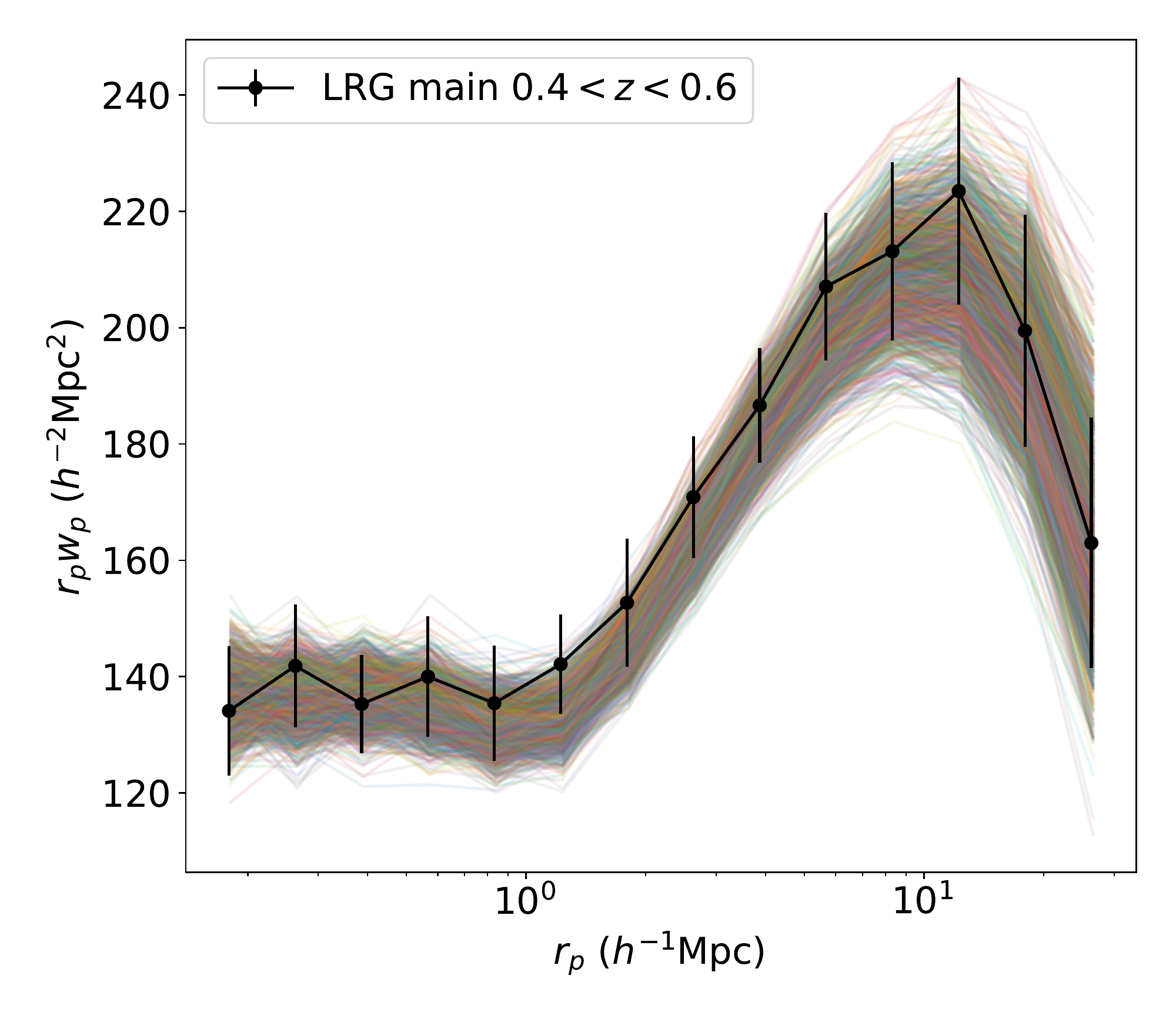}
    \vspace{-0.3cm}
    \caption{The projected auto-correlation function $w_p$ of the 1800 boxes after tuning vanilla HOD parameters to match the clustering of One-Percent Survey LRGs in $0.4 < z < 0.6$. See section~\ref{sec:cov} for details.}
    \label{fig:wpsmall}
\end{figure}

Figure~\ref{fig:wpsmall} serves as a visualisation of the 1800 realisations after tuning to match the observed \xirppi, where we overlay the projected auto-correlation functions of the 1800 boxes on the observation. We see that the mock realisations do well in producing the observed clustering, and the spread in the mock clustering is consistent in trend with the data jackknife errorbars. 

\begin{figure}
    \hspace{-0.7cm}
    \includegraphics[width=0.53\textwidth]{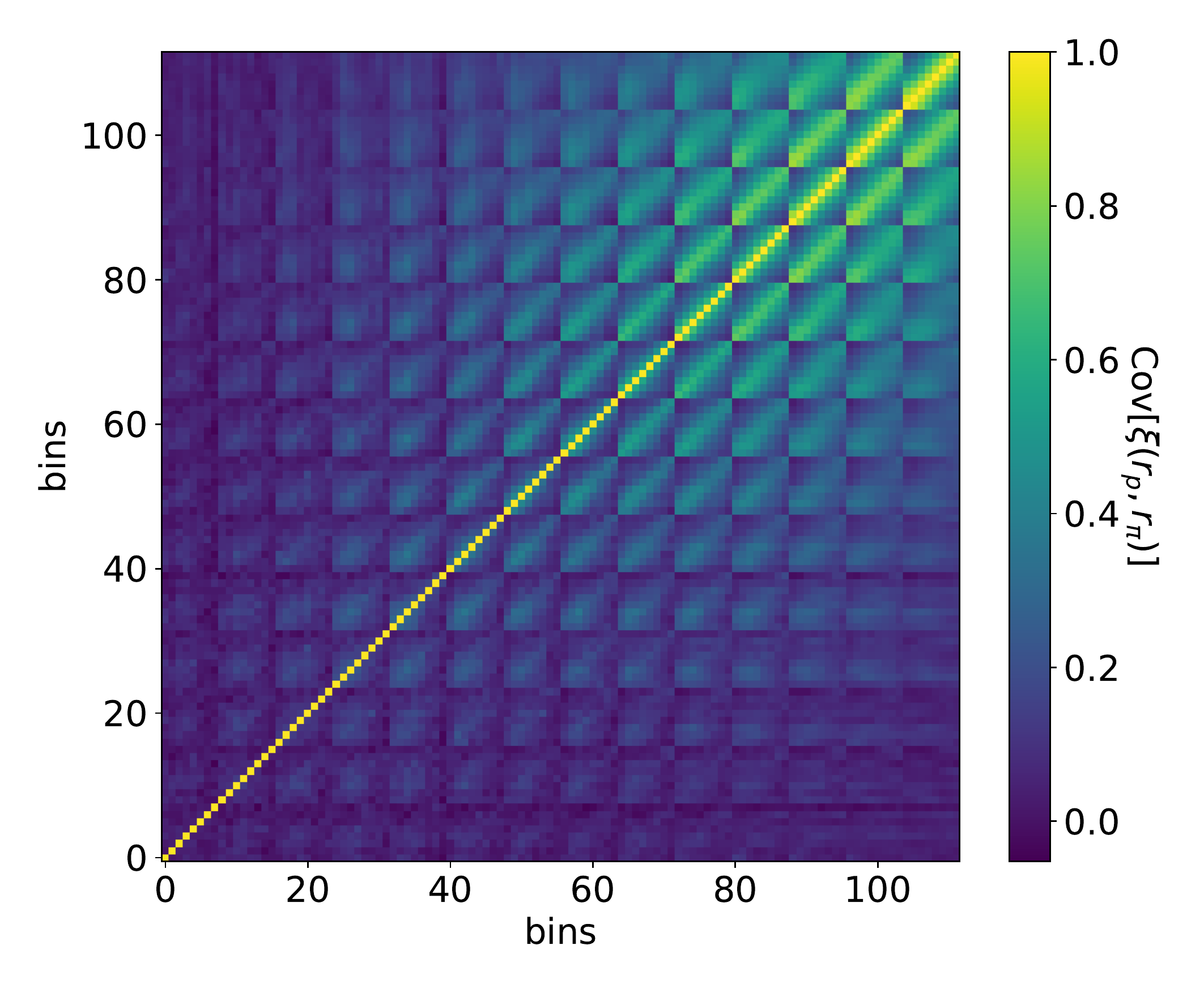}
    \vspace{-0.7cm}
    \caption{The correlation matrix of One-Percent Survey LRG \xirppi in $0.4 < z < 0.6$, generated from the 1800 boxes after tuning to match the observed clustering. The 2D bins of \xirppi are collapsed in this representation as a column-wise stack (the bin number is strictly increasing in $r_p$ and periodic in $r_\pi$)}
    \label{fig:corr}
\end{figure}

Figure~\ref{fig:corr} shows the resulting correlation matrix computed from the 1800 mocks for the LRG \xirppi in $0.4 < z < 0.6$. The 2D bins of \xirppi are collapsed in this representation as a column-wise stack (the bin number is strictly increasing in $r_p$ and periodic in $r_\pi$). We see that the off-diagonal terms are determined with high signal-to-noise, a result of the large number of realisations and the large volume available with the 1800 boxes. The large-scale bins are correlated as they become sample variance dominated by large-scale structure, whereas the small-scale bins are largely independent, as they are dominated by the shot noise of galaxy occupation. 

We similarly generate mock covariance matrices for the LRG sample in $0.6 < z < 0.8$ and the QSO sample. We omit those plots for brevity. The covariance matrix for the LRG sample  in $0.6 < z < 0.8$ share essentially the same structure as the lower redshift LRG sample, with shot noise dominating the smallest scales and sample variance becoming significant at larger scales. For QSOs, all scales are dominated by shot noise and the covariance matrix is essentially diagonal. Throughout the rest of this paper, we use these set of mock covariance matrices for model comparison and posterior sampling. 

Having defined the likelihood function, we use optimisation routines and posterior samplers to evaluate the best-fits and posterior constraints, respectively. Finally, we note that a correction term is applied to correct for the finite number of independent realisations used to calculate the covariance matrix \citep{2007Hartlap}. Due to a large number of mock realisations ($\sim$1800), the correction factor is small but not negligible ($\sim 6\%$). 

\section{LRG HOD Results}
\label{sec:results}
In this section, we present the results of the One-Percent Survey LRG HOD analysis by deriving the HOD best-fits, testing possible model extensions, and presenting the posterior constraints. 

\subsection{LRG at $z < 0.8$}

\begin{figure}
    \hspace{-0.3cm}
    \includegraphics[width=0.5\textwidth]{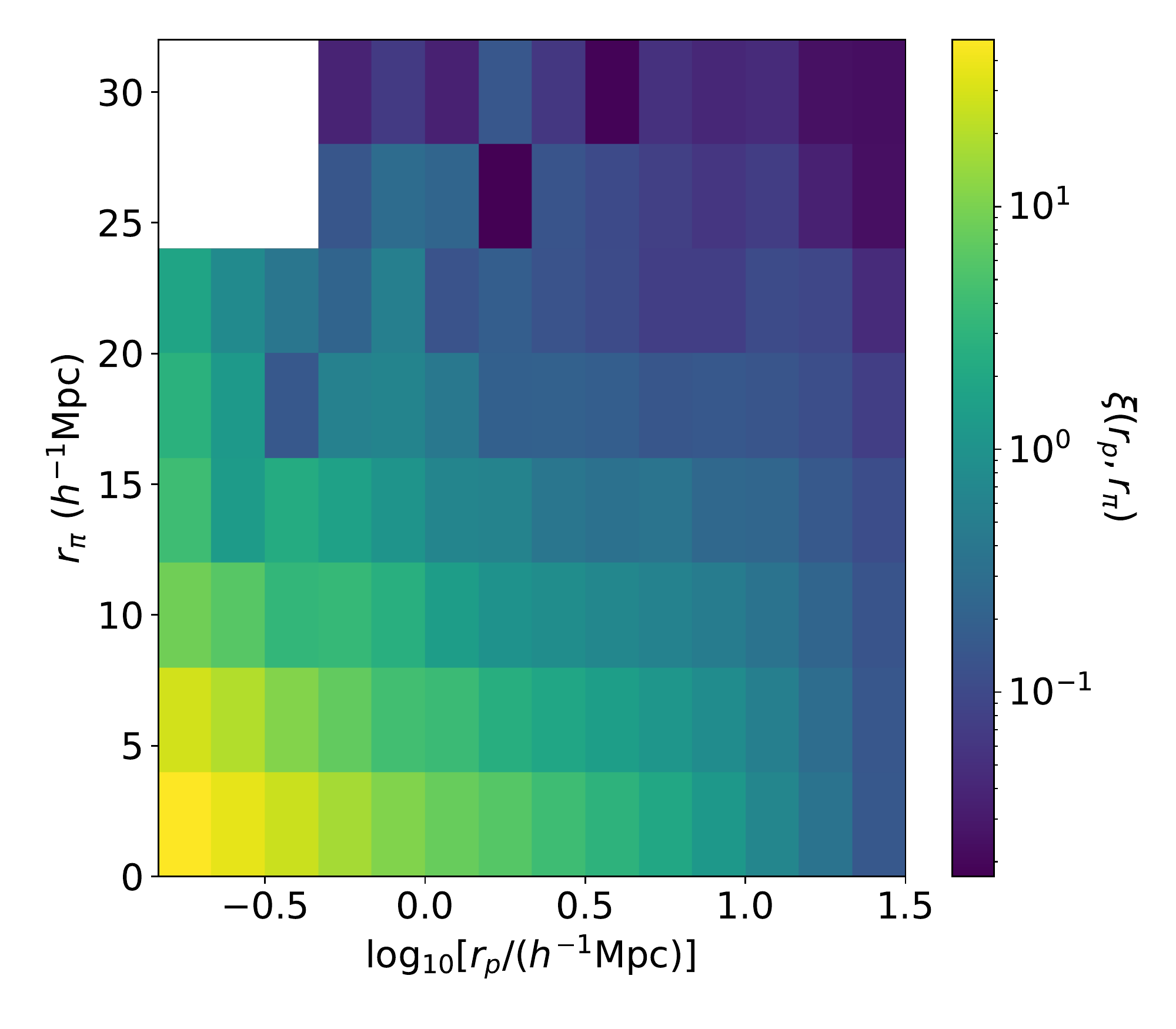}
    \vspace{-0.3cm}
    \caption{The One-Percent Survey LRG \xirppi in $0.4 < z < 0.6$. $r_p$ and $r_\pi$ denote the transverse and LoS separation of the galaxies in comoving units. For this analysis, we only utilise the transverse scales 0.15--32$\ h^{-1}$Mpc. The white section corresponds to negative values, which do not show up on the log scale. }
    \label{fig:xi2d}
\end{figure}

We first examine the LRG main sample at $z<0.8$, where the number density remains relatively constant. This is also the LRG sample that will be used for DESI Y1 fiducial cosmology analyses. We analyze this sample in two separate redshift bins: $0.4 < z < 0.6$ and $0.6 < z < 0.8$, using the $z=0.5$ and $z=0.8$ snapshots, respectively. 
%\sout{We analyse the LRG main sample in two different redshift bins $0.4 < z < 0.6$ and $0.6 < z < 0.8$. Note that the reason for limiting to $z < 0.8$ is to use only the redshift range where the $n(z)$ is approximately constant.} 
We target the $\xi(r_\mathrm{p},r_\pi)+n(z)$ data vector in each bin and incorporate the full covariance matrix built on mocks. Figure~\ref{fig:xi2d} shows the LRG \xirppi data vector in $0.4 < z < 0.6$. We omit the visualisations for the other redshift bin for brevity. We only utilise the transverse scales 0.15--32$\ h^{-1}$Mpc and LoS separation from 0 to $32\ $\hmpc.

% In each redshift bin, we first run one optimization chain with a vanilla plus velocity bias model on the \xirppi data vector with data jackknife errors, assuming no off-diagonal covariances. We find good fits for both redshift bins. We list the best-fit parameters in the ``diagonal-only'' columns of Table~\ref{tab:LRG_fit1}. With the $0.4 < z < 0.6$, we get best-fit $\chi^2 = 79.6$ with 105 degrees of freedom, and for $0.6 < z < 0.8$, we get best-fit $\chi^2 = 91.6$ with 105 degrees of freedom. The rows of Table~\ref{tab:LRG_fit1} list not only the best-fit HOD parameters, but also inferred parameters such satellite fraction $f_\mathrm{sat}$ and average logrithmic halo mass $\log_{10}\overline{M}_h$. 

% As described in section~\ref{sec:cov}, we first generate the mock-based covariance matrices by running HOD fits with just the data jackknife errors. 
% We re-normalize these covariance matrices with the data jackknife errors by taking the mock-based correlation matrices and multiplying the $i, j$-th term with $\sigma_i\sigma_j$, where $\sigma_i$ is the data jackknife error in the $i$-th bin. We use these rescaled covariance matrices as the fiducial covariances in ensuring fits. 
We first test for potential extensions to the 8-parameter baseline HOD model (5-parameter vanilla HOD plus velocity bias plus incompleteness). Specifically, we run optimisations with extended models that include either galaxy assembly bias or satellite radial profile parameter in addition to the baseline parameters. We use a global optimisation routine called Covariance matrix adaptation evolution strategy (CMA-ES) with 400 random walkers. We compute the model Akaike Information Criterion (AIC) scores from the best-fit $\chi^2$ and summarise the results in Table~\ref{tab:AIC}. The AIC scores essentially calculates the best-fit $\chi^2$ but compensating for the number of parameters. Models with lower AIC scores are preferred by the data, and a $\Delta \mathrm{AIC} =1$ roughly corresponds to 1$\sigma$ significance. 

Our tests show no evidence for either flavors of galaxy assembly bias or a satellite radial profile parameter. We conclude that the current data vectors favor the baseline model, and we will conduct the rest of this analysis with just the baseline HOD model. However, \cite{2021bYuan} found significant evidence for galaxy assembly bias in a similar HOD analysis of BOSS CMASS LRGs. This discrepancy is explained by the significantly larger sample size in CMASS ($\sim$600,000). The factor 10 decrease in sample size translates to a factor 3 increase in statistical error, which in turn downgrades an assembly bias signal as detected in \cite{2021bYuan} to less than $1\sigma$ significance. 
\begin{table}
\centering
\begin{tabular}{l || lclr}
\hline
\hline
LRG & \multicolumn{2}{c}{$0.4 < z < 0.6$} & \multicolumn{2}{c}{$0.6 < z <0.8$}\\
\hline
Model & $\chi^2$/d.o.f. & AIC & $\chi^2$/d.o.f. & AIC \\
\hline
Baseline      & 1.03 & 130 & 0.99& 125\\
Baseline+$A$  & 1.05 & 133 & 0.98& 125 \\
Baseline+$B$  & 1.05 & 133 & 0.99& 126\\
Baseline+$s$  & 1.06 & 131 & 1.01& 126\\
\hline
\end{tabular}
\caption{Comparing LRG HOD model extensions in both redshift bins. The baseline model refers to the standard vanilla 5-parameter model plus velocity bias and incompleteness. $A$ refers to galaxy assembly bias parameterised in terms of halo concentration. $B$ refers to galaxy assembly bias parameterised in terms of the local environment. $s$ refers to 1-halo satellite profile modulations. The AIC scores suggest that none of the extended models are preferred over the baseline model.}
\label{tab:AIC}
\end{table}

\begin{table}
    \centering
    \begin{tabular}{l||cc|cc}
        \hline
        \hline
        \multirow{2}{*}{Params} &\multicolumn{2}{c|}{LRG}&\multicolumn{2}{c}{QSO}\\
        \cline{2-5}
                                &prior                  &bounds     &prior      &bounds\\
        \hline
        $\log M_\mathrm{cut}$   &$\mathcal{N}(13.0,1.0)$&[12,13.8]  &$\mathcal{N}(12.7, 1)$&[11.2, 14.0]\\
        $\log M_1$              &$\mathcal{N}(14.0,1.0)$&[12.5,15.5]&$\mathcal{N}(15.0, 1)$&[12.0, 16.0]\\
        $\sigma$                &$\mathcal{N}(0.5,0.5)$ &[0.0,3.0]  &$\mathcal{N}(0.5, 0.5)$&[0.0, 3.0]\\
        $\alpha$                &$\mathcal{N}(1.0,0.5)$ &[0.0,2.0]  &$\mathcal{N}(1.0, 0.5)$&[0.3, 2.0]\\
        $\kappa$                &$\mathcal{N}(0.5,0.5)$ &[0.0,10.0] &$\mathcal{N}(0.5, 0.5)$&[0.3, 3.0]\\
%        \hline
        % $f_\mathrm{ic}$         &&&&\\
%        \hline
        $\alpha_\mathrm{c}$     &$\mathcal{N}(0.4,0.4)$&[0.0, 1.0]&$\mathcal{N}(1.5,1.0)$&[0.0, 2.0]\\
        $\alpha_\mathrm{s}$     &$\mathcal{N}(0.8,0.4)$&[0.0, 2.0]&$\mathcal{N}(0.2, 1.0)$&[0.0, 2.0]\\
        \hline        
    \end{tabular}
    \caption{Priors used for LRG and QSO HOD fits. We use broad Gaussian priors on all parameters. We also quote the bounds we impose in addition to the Gaussian priors. Units of mass are in $h^{-1}M_\odot$.
    }
    \label{tab:priors}
\end{table}
\begin{table*}
    \centering
    {\renewcommand{\arraystretch}{1.5}
    %\resizebox{\textwidth}{!}{%
    \begin{tabular}{l||c|c|c|c|c|c}
        \hline
        \hline
        Tracer&\multicolumn{2}{c|}{$\mathrm{LRG}\;\;0.4<z<0.6$}&\multicolumn{2}{c|}{$\mathrm{LRG}\;\;0.6<z<0.8$}&\multicolumn{2}{c}{$\mathrm{QSO}\;\;0.8<z<2.1$}\\[2pt]
        \hline
        Model&
        $\mathrm{Zheng07}+f_\mathrm{ic}$&\makecell[cl]{$\mathrm{Zheng07}+f_\mathrm{ic}$\\$\;\;+\alpha_\mathrm{c}+\alpha_\mathrm{s}$}&
        $\mathrm{Zheng07}+f_\mathrm{ic}$&\makecell[cl]{$\mathrm{Zheng07}+f_\mathrm{ic}$\\$\;\;+\alpha_\mathrm{c}+\alpha_\mathrm{s}$}&
        $\mathrm{Zheng07}+f_\mathrm{ic}$&\makecell[cl]{$\mathrm{Zheng07}+f_\mathrm{ic}$\\$\;\;+\alpha_\mathrm{c}+\alpha_\mathrm{s}$}\\
        \hline
        Data&$w_\mathrm{p}+n_z$&\xirppi$+n_z$&$w_\mathrm{p}+n_z$&\xirppi$+n_z$&$w_\mathrm{p}+n_z$&\xirppi$+n_z$\\[2pt]
        
        \hline
        $\log M_\mathrm{cut}$&
        12.89$^{+0.11}_{-0.09}$&
        12.79$^{+0.15}_{-0.07}$&
        12.78$_{-0.08}^{+0.10}$&
        12.64$^{+0.17}_{-0.05}$&
        12.67$_{-0.36}^{+0.71}$&
        12.2$^{+0.6}_{-0.1}$\\
        
        $\log M_1$&
        14.08$^{+0.10}_{-0.10}$&
        13.88$^{+0.11}_{-0.11}$&
        13.94$_{-0.11}^{+0.14}$&
        13.71$^{+0.07}_{-0.07}$&
        15.00$_{-0.64}^{+0.62}$&
        14.7$^{+0.6}_{-0.6}$\\

        $\sigma$&
        0.27$^{+0.17}_{-0.17}$&
        0.21$^{+0.11}_{-0.10}$&
        0.23$_{-0.15}^{+0.14}$&
        0.09$^{+0.09}_{-0.05}$&
        0.58$_{-0.35}^{+0.37}$&
        0.12$^{+0.28}_{-0.06}$\\
        
        $\alpha$&
        1.20$^{+0.15}_{-0.19}$&
        1.07$^{+0.13}_{-0.16}$&
        1.07$_{-0.21}^{+0.16}$&
        1.18$^{+0.08}_{-0.13}$&
        1.09$_{-0.37}^{+0.43}$&
        0.8$^{+0.4}_{-0.2}$\\
        
        $\kappa$&
        0.65$^{+0.45}_{-0.39}$&
        1.4$^{+0.6}_{-0.5}$&
        0.55$_{-0.34}^{+0.42}$&
        0.6$^{+0.4}_{-0.2}$&
        0.74$_{-0.29}^{+0.41}$&
        0.6$^{+0.8}_{-0.2}$\\[2pt]

        \hline
        
        $\alpha_c$&
        -&0.33$^{+0.05}_{-0.07}$&
        -&0.19$^{+0.06}_{-0.09}$&
        -&1.54$^{+0.17}_{-0.08}$\\
        
        $\alpha_s$&
        -&0.80$^{+0.07}_{-0.07}$&
        -&0.95$^{+0.07}_{-0.06}$&
        -&0.6$^{+0.6}_{-0.3}$\\[2pt]

        \hline
        
        $f_\mathrm{ic}$&
        0.92$^{+0.08}_{-0.17}$&
        0.70$^{+0.15}_{-0.09}$&
        0.89$_{-0.14}^{+0.11}$&
        0.62$^{+0.07}_{-0.06}$&
        0.041$_{-0.016}^{+0.066}$&
        0.019$^{+0.029}_{-0.004}$\\
        
        $f_\mathrm{sat}$&
        0.089$_{-0.010}^{+0.013}$&
        0.106$^{+0.011}_{-0.012}$&
        0.104$_{-0.010}^{+0.013}$&
        0.136$^{+0.011}_{-0.010}$&
        0.05$_{-0.05}^{+0.26}$&
        0.03$^{+0.08}_{-0.02}$\\

        $\log\overline{M}_\mathrm{h}$&
        13.42$_{-0.02}^{+0.02}$&
        13.40$^{+0.02}_{-0.02}$&
        13.26$_{-0.02}^{+0.02}$&
        13.24$_{-0.02}^{+0.02}$&
        12.74$_{-0.05}^{+0.05}$&
        12.65$^{+0.09}_{-0.04}$\\

        $b_\mathrm{lin}$&
        $1.94_{-0.04}^{+0.04}$&
        $1.93_{-0.04}^{+0.06}$&
        $2.11_{-0.04}^{+0.03}$&
        $2.08_{-0.03}^{+0.03}$&
        $2.56_{-0.10}^{+0.22}$&
        $2.63_{-0.26}^{+0.37}$\\[2pt]

        \hline
        
        $\chi^2/\mathrm{d.o.f}$&
        4.5/(14-5)&108/(112-7)&19.6/(14-5)&104/(112-7)&16.0/(14-5)&101/(112-7)\\

        \hline
    \end{tabular}%
    %}
    }
    \caption{LRG and QSO marginalized posteriors, with different models and different measurements. The error bars are $1\sigma$ uncertainties. We also display several derived parameters, specifically the marginalized satellite fraction $f_\mathrm{sat}$, the sample completeness $f_\mathrm{ic}$, the average halo mass per galaxy $\log\overline{M}_\mathrm{h}$, and the linear bias$b_\mathrm{lin}$. Units of mass are given in $h^{-1}M_\odot$.
    }
    \label{tab:LRG_fit1}
\end{table*}

\begin{figure*}
    \hspace{-0.7cm}
    \includegraphics[width=1\textwidth]{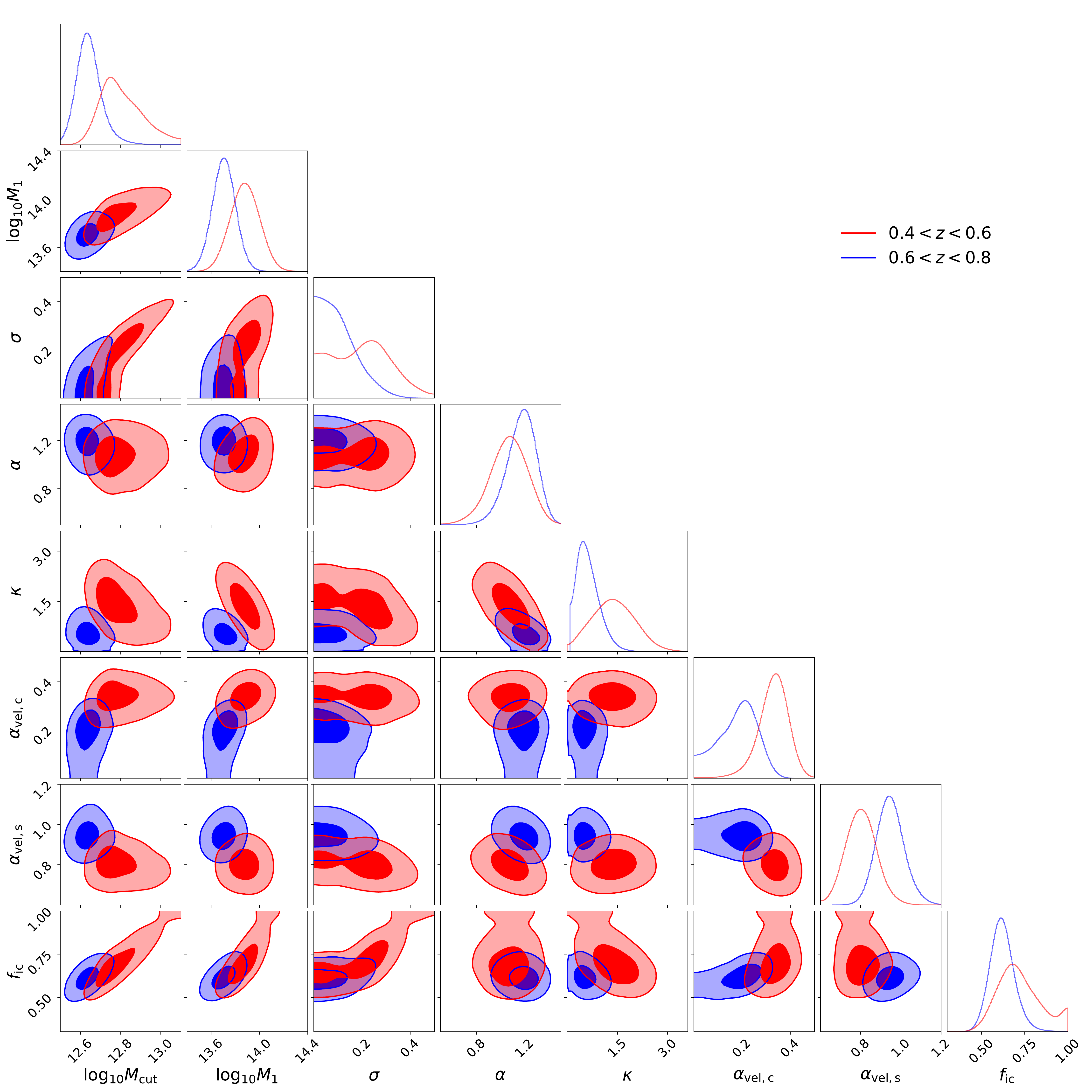}
    \vspace{-0.2cm}
    \caption{The DESI One-Percent Survey LRG HOD posterior. The red and blue contours correspond to the marginalised posteriors for LRGs in the $0.4 < z < 0.6$ and $0.6 < z < 0.8$ bins, respectively. We are showing only the 1 and 2$\sigma$ contours for clarity.}
    \label{fig:corner_lrg}
\end{figure*}

\begin{figure*}
    \begin{subfigure}[b]{0.45\textwidth}
         \hspace{-0.7cm}
         \centering
         \includegraphics[width=\textwidth]{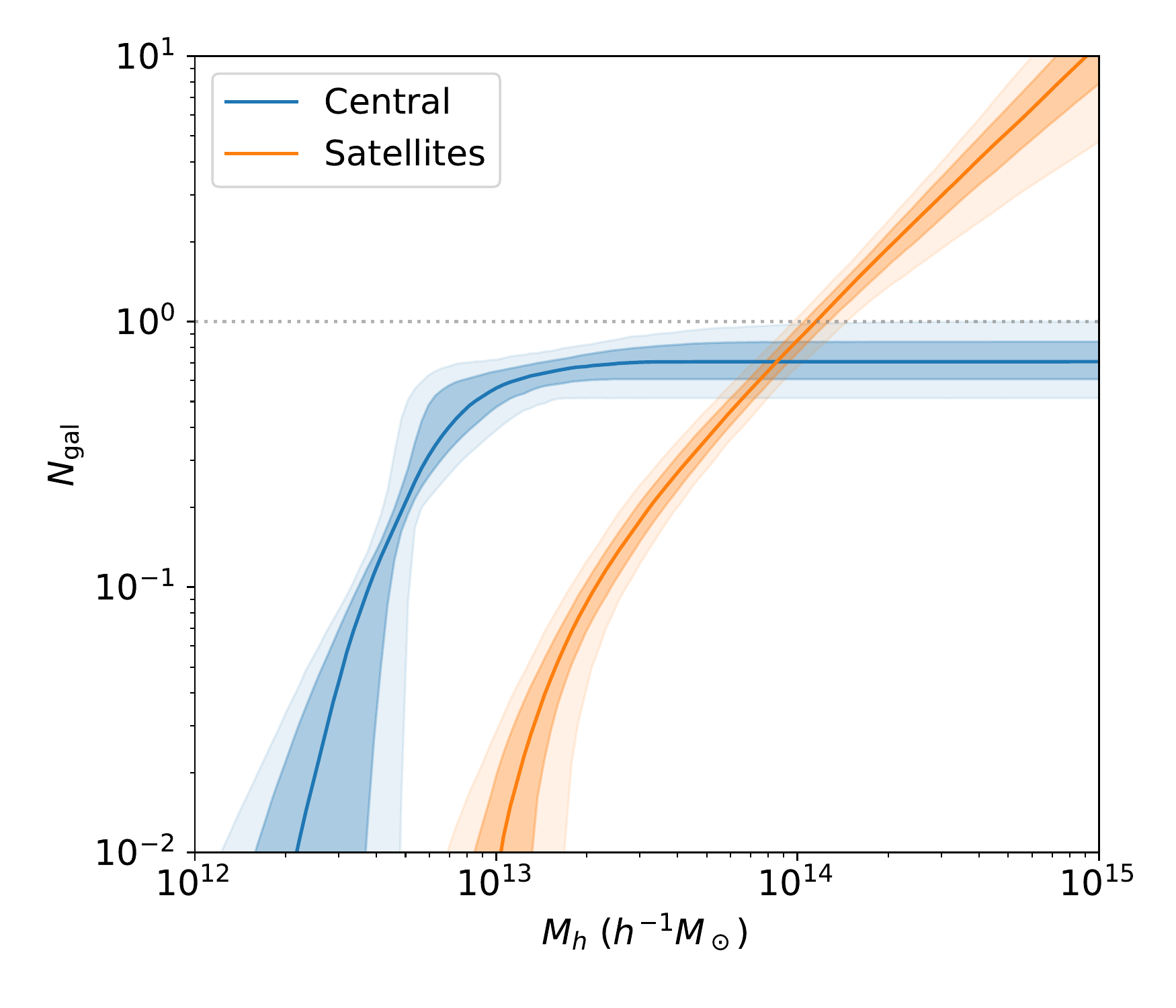}
        \vspace{-0.4cm}
         \caption{LRG $0.4 < z < 0.6$}
         \label{fig:hod5}
     \end{subfigure}
    \begin{subfigure}[b]{0.45\textwidth}
         \hspace{-0.7cm}
             \centering
         \includegraphics[width=\textwidth]{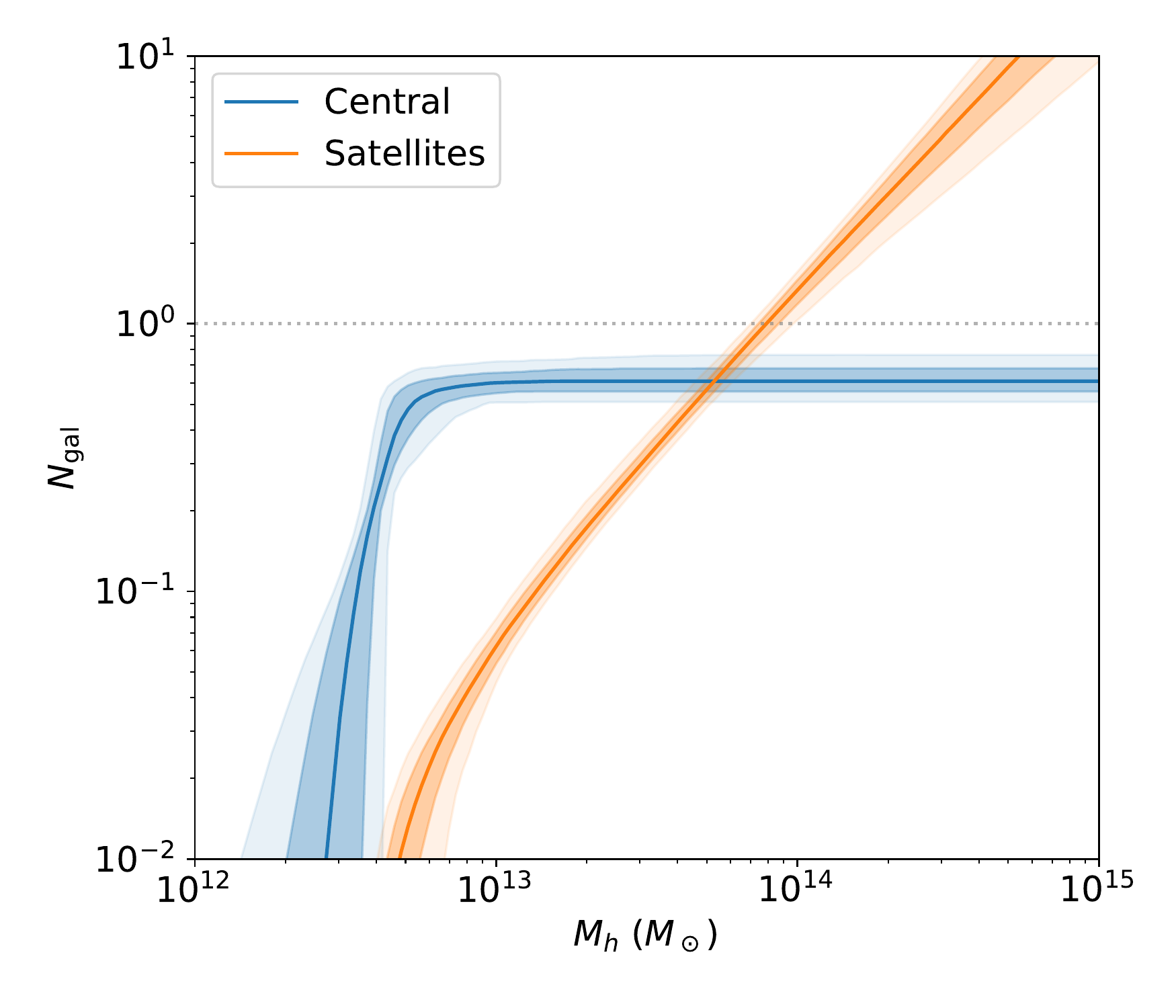}
        \vspace{-0.4cm}
         \caption{LRG $0.6 < z < 0.8$}
         \label{fig:hod8}
     \end{subfigure}
    \caption{The LRG HOD best-fit the posterior. The shaded regions correspond to 1 and 2$\sigma$ posteriors ($68\%$ and $95\%$ intervals centered around the median prediction). The horizontal dotted line denotes $N_\mathrm{gal} = 1$. }
    \label{fig:hod}
\end{figure*}

% For the dynesty runs, we find that the parameter $\sigma$ is very poorly constrained for both LRG samples due to a strong degeneracy with $\log M_\mathrm{cut}$. Thus, we fix $\sigma$ to be 0.15 in both cases to achieve better convergence on other HOD parameters. The value 0.15 is arbitrary but chosen to be small because we see a very mild preference for small $\sigma$ value in its marginalised posterior. The choice of $\sigma$ has essentially no effect on the best-fit $\chi^2$. \sandy{But does it affect logMcut} These fits are also not good. Now we try gaussian priors centered on the optimised fits.
Next, we obtain the parameter posteriors for the baseline model in each redshift bin. To sample the HOD posterior, we use the efficient \textsc{dynesty} nested sampler \citep{2018Speagle, 2019Speagle}. Again, we use \xirppi as our primary data vector in each redshift bin and include the full covariance matrix in the likelihood evaluation. We incorporate broad multivariate Gaussian priors as quoted in Table~\ref{tab:priors}. We specifically test our choice of Gaussian priors will not significantly bias our main results. (see App.~\ref{app:prior} for details) We also impose bounds to limit the range explored by the sampler (also documented in Table~\ref{tab:priors}). The resulting marginalised posteriors are presented in Figure~\ref{fig:corner_lrg} and summarised in Table~\ref{tab:LRG_fit1}. We visualise the corresponding HOD posteriors in Figure~\ref{fig:hod}, where the shaded bands denote the 1 and 2$\sigma$ constraints ($68\%$ and $95\%$ intervals). 

% show full parameter posteriors, possibly stacking the two redshift bins. 

% discuss the corner pltos
Figure~\ref{fig:corner_lrg} shows several interesting degeneracies. Perhaps the most prominent degeneracy is between parameter $\log M_\mathrm{cut}$ and $\sigma$. Both parameters control the occupation distribution of the centrals, which for LRGs translates to controlling the clustering amplitude of the 2-halo term on large scales. $\log M_\mathrm{cut}$ controls the mass scale whereas $\sigma$ controls the slope of the $N_\mathrm{cent}$ turn-on. Thus, it makes sense that the two are somewhat degenerate as either a slower turn on (larger $\sigma$) or a lower mass scale would decrease the mean bias of the sample. 

Similarly, we see a degeneracy between the completeness parameter $f_\mathrm{ic}$ and the central occupation parameters $\log M_\mathrm{cut}$ and $\sigma$. This is due to the constraints on the average density of the galaxy sample. Lower typical halo masses for the galaxies would mean a lower completeness. There is also an interesting degeneracy between $\log M_\mathrm{cut}$ and $\log M_1$, which has been previously discussed in \citet{2020Avila}. This is interesting because $\log M_1$ controls the halo mass of the satellite galaxies. Thus, a degeneracy between these two parameters suggest a strong constraint on the satellite fraction, and by extension a strong constraint on the relative amplitude of the 1-halo clustering and 2-halo clustering. 

% Comparing the best-fit parameters of the diagonal-only run and the full covariance matrix run in Table~\ref{tab:LRG_fit1}, we see that the parameters are largely consistent between the two runs, suggesting that the additional of off-diagonal powers in the covariance matrices did not significantly bias the constraints. At the bottom of the table, we also list two inferred parameters, satellite fraction $f_\mathrm{sat}$ and average halo mass $\log_{10}\overline{M}_h$. 

The inferred satellite fraction is $11\pm 1\%$ for LRGs in $0.4 < z < 0.6$ and $14\pm 1\%$ in $0.6 < z < 0.8$, consistent with $11\%$ inferred for the CMASS sample in $0.45 < z < 0.6$ \citep{2021bYuan} and $13\pm 3\%$ inferred with eBOSS LRGs between $0.6 < z < 0.9$ \citep{2017Zhai}. The mean halo mass of the LRGs is strongly constrained. We find \smash{$\log_{10}\overline{M}_h=13.40^{+0.02}_{-0.02}$} for $0.4 < z < 0.6$ and \smash{$\log_{10}\overline{M}_h=13.24^{+0.02}_{-0.02}$} for $0.6 < z < 0.8$. In comparison, \cite{2021Yuan} found $\log_{10}\overline{M}_h = 13.60$ for the CMASS sample, and \cite{2017Zhai} found $\log_{10}\overline{M}_h = 13.4$ for the higher redshift eBOSS sample. Both values are higher than the average halo mass inferred for DESI One-Percent Survey LRGs. This is expected as the DESI sample is fainter and higher number density, thus occupying less massive halos. Our results also compare well with earlier results from CMASS DR10, which reported a 9-10$\%$ satellite fraction for a red luminosity-limited sample with half the DESI density \citep{2014Guo}. The difference in average halo mass between the two redshift bins can simply be attributed to halo growth at fixed density. 

The linear bias factor is also calculated in our study by comparing the real-space clustering amplitudes of galaxies with predictions made by the linear theory, specifically
\begin{equation}
b_\mathrm{lin}=(\xi_\mathrm{gals}/\xi_\mathrm{lin})^{1/2},
\end{equation}
where $\xi_\mathrm{gal}$ denotes the real-space two-point correlation functions of galaxies, while $\xi_\mathrm{lin}$ denotes the theoretical linear matter correlation function at the mean redshift in the respective redshift bin. This correlation function is measured with \texttt{CLASS} \citep{2011arXiv1104.2932L}. For each sample of the MCMC chain, $b_\mathrm{lin}$ is calculated for a uniform subsample of the full posterior, which we summarise with mean and standard deviation. The LRG linear bias is \smash{$1.93^{+0.06}_{-0.04}$} in the redshift range of $0.4<z<0.6$ and $2.08\pm0.03$ for LRGs in $0.6<z<0.8$. We present a more detailed description of redshift evolution in section~\ref{subsec:hodz}. %\sandy{Ashley: Can you say anything about whether the difference in halo mass is basically what is expected from the mass growth of halos between the two redshifts? Is there anything interesting to say abou the change in the satellite fraction? I see now a later paragraph discusses the trends more. It might be helpful to put that paragraph closer to where the results are reported (and I think you can get the exact expected mass growth for a constant number density of halos, at least given some mass function?)}

% eboss 13±3 \% and a mean halo mass of 2.5×1013h−1M⊙

The velocity bias constraints are also mostly consistent with BOSS and eBOSS studies. For the $0.4 < z < 0.6$ bin, we find significant central velocity bias at \smash{$33^{+5}_{-7}\%$}, indicating that the peculiar velocity of the central galaxies relative to the central subhalos are approximately $30\%$ of the halo velocity dispersion. This is qualitatively consistent with previous studies that also found significant central velocity bias, but somewhat larger in amplitude than the CMASS constraints at $22\pm 2\%$ \citep{2022Yuan, 2015aGuo}. We also find negative satellite velocity bias $80\pm 7\%$, indicating that the velocity dispersion of satellite galaxies within halos are $20\%$ less than that of the halo particles at the same radii. This is consistent with \cite{2015aGuo}, who found $86\%$ in CMASS, whereas \cite{2022Yuan} found less significant velocity bias at $98\%$. In the higher redshift bin $0.6 < z < 0.8$, we find a central velocity bias of \smash{$19^{+6}_{-9}\%$} and a satellite velocity bias \smash{$95^{+7}_{-6}\%$}. We do not have eBOSS constraints to compare against, but we can check with simulated DESI samples presented in \cite{2022Yuan}, where we applied DESI photometric selection to \textsc{IllustrisTNG} galaxies \citep{2018Pillepich, 2018Springel, 2018Nelson}. There, we found that the mock DESI LRG sample at $z = 0.8$ has a velocity bias of $\alpha_c = 0.14$ and $\alpha_s = 0.92$, consistent with our constraints here. 
%\sandy{Ashley: Is there any reason to believe the central velocity bias should change significantly with redshift, or is it more likely one would want the mean of the two measurements as the best estimate? (It might be hard to address such a question in the results section, but it is something I wonder while reading.)

We also run the equivalent analyses on the projected 2PCF $w_\mathrm{p}$, where we follow the same procedure as for \xirppi, but we exclude the two velocity bias parameters from the HOD model. We also employ a tabulation scheme to accelerate the HOD forward model calculation for these $w_p$ fits (details in App.~\ref{app:tab}). We include the $w_p$ marginalised constraints in Table~\ref{tab:LRG_fit1}. However, for brevity, we skip their visualisation. Comparing the marginalised posteriors in Table~\ref{tab:LRG_fit1}, the $w_p$ results are consistent with the \xirppi\ results. In terms of inferred quantities, the two data vectors also yield mostly consistent results. The only minor discrepancy between the two fits is that the $w_p$ fits favor larger $M_\mathrm{cut}$ and $M_1$ parameters, and as a result a larger $f_\mathrm{ic}$. However, this is compensated by differences in the $\sigma$ constraints, resulting in essentially identical mean halo mass constraints. In other words, both data vectors place strong constraints on the mean halo mass of the galaxies, but neither breaks the $M_\mathrm{cut}$ vs. $\sigma$ degeneracy and favor slightly different loci along this degeneracy.

Comparing the posterior means in the two redshift bins, we see several interesting trends. The average halo mass of the LRG sample increases over time. It is expected given that halos accrete mass over time, and a fixed density sample at lower redshift would occupy the more massive halos. The satellite fraction also decreases over time. This trend is not as significant but might be interpreted as a result of galaxy mergers. We discuss redshift evolution in more detail in section~\ref{subsec:hodz}. 
% show HOD plot

\begin{figure*}
    \begin{subfigure}[b]{\textwidth}
         \hspace{-0.7cm}
         \centering
         \includegraphics[width=\textwidth]{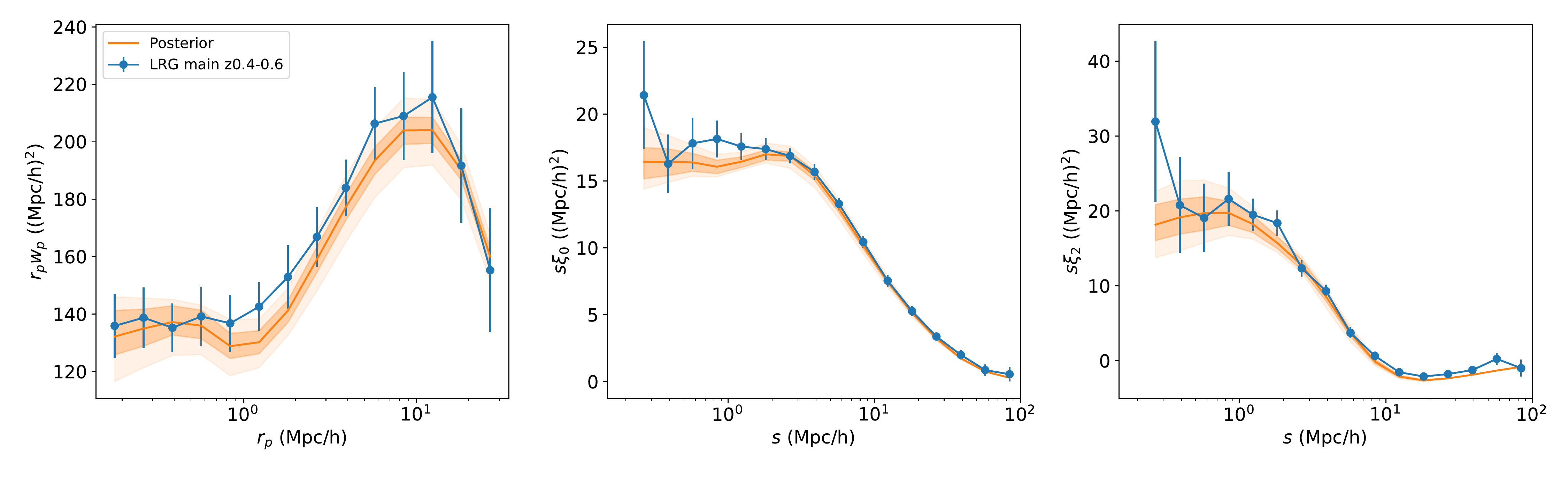}
        \vspace{-0.4cm}
         \caption{LRG $0.4 < z < 0.6$}
         \label{fig:mulitpole5}
     \end{subfigure}
    \begin{subfigure}[b]{\textwidth}
         \hspace{-0.7cm}
             \centering
         \includegraphics[width=\textwidth]{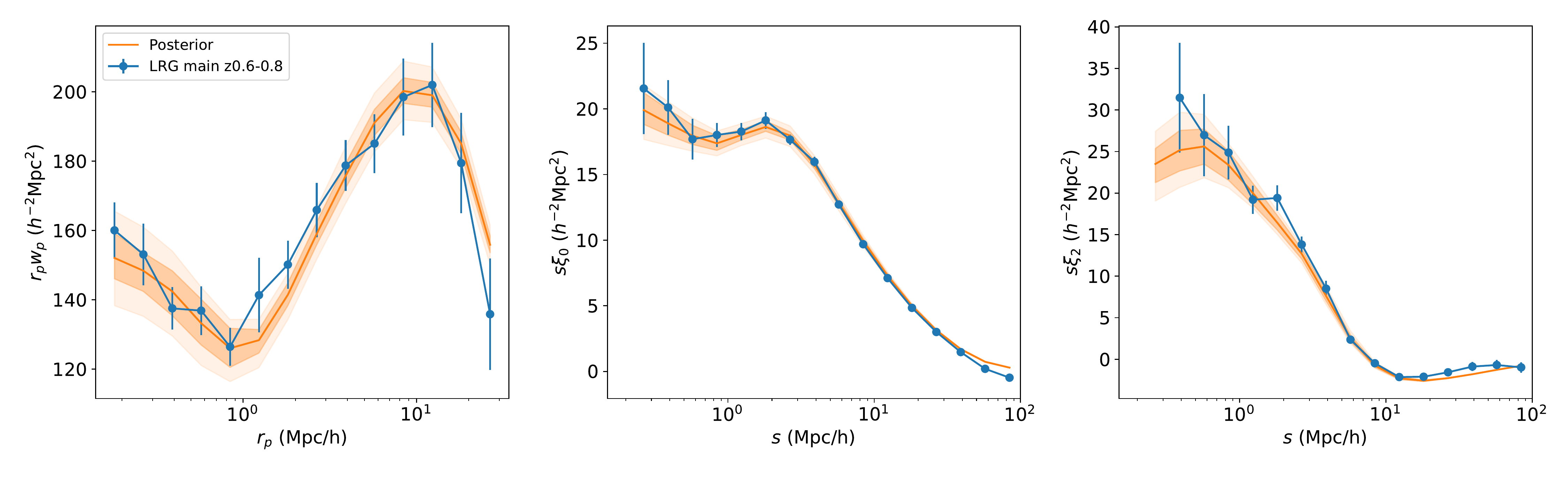}
        \vspace{-0.4cm}
         \caption{LRG $0.6 < z < 0.8$}
         \label{fig:multipole8}
     \end{subfigure}
    \caption{The LRG $w_p$ and multipoles posterior predicatives compared to the data. The blue lines correspond to the One-Percent Survey measurement with jackknife error bars. The solid orange line denote the posterior mean. The orange shaded regions correspond to the 1 and 2$\sigma$ full posterior using the full mock covariance matrix.}
    \label{fig:multipoles}
\end{figure*}

% show clustering predictions
Finally, Figure~\ref{fig:multipoles} showcases the predicted distribution of the 2PCF from the \xirppi posteriors. For this visualisation, we choose the $w_p+\xi_0+\xi_2$ projections of the redshift-space 2PCF because it is difficult to visualise comparisons of the 2D \xirppi function. $w_p$ is the projected 2PCF, whereas $\xi_{0,2}$ are the monopole and quadruple moments of the redshift-space 2PCF. The blue curves represent the One-Percent Survey measurements, with jackknife errorbars. The orange shaded regions denote the 1 and 2$\sigma$ posterior constraints. The solid orange line showcases the posterior mean. Again, we see that the best-fit models are consistent with the observed clustering well in both redshift bins. However, in the $w_p$ comparisons, the model predicts a larger amplitude than the data at the 1-halo and 2-halo transition regime ($r_p\sim 1h^{-1}$Mpc) in both redshift bins. This is not a significant discrepancy with the current sample size but potentially points to limitations of halo boundary definitions and possibly environment-dependent galaxy occupation.

\subsection{LRG at $z > 0.8$}
\label{subsec:lrghighz}

\begin{table}
    \centering
    {\renewcommand{\arraystretch}{1.5}
    %\resizebox{0.9\linewidth}{!}{%
    \begin{tabular}{l||c|c}
        \hline
        \hline
        \multirow{2}{*}{Params} &\multicolumn{2}{c}{LRG $0.8<z<1.1$}\\
        \cline{2-3}
                                &$0.8<z<0.95$               &$0.95<z<1.1$\\                  
        \hline
        $\log M_\mathrm{cut}$   &12.89$_{-0.13}^{+0.12}$    &12.68$_{-0.26}^{+0.38}$\\
        $\log M_1$              &13.96$_{-0.14}^{+0.15}$    &13.60$_{-0.29}^{+0.47}$\\
        $\sigma$                &0.37$_{-0.18}^{+0.13}$     &0.53$_{-0.29}^{+0.25}$\\
        $\alpha$                &0.91$_{-0.22}^{+0.18}$     &0.72$_{-0.34}^{+0.31}$\\
        $\kappa$                &0.74$_{-0.42}^{+0.46}$     &0.51$_{-0.33}^{+0.43}$\\[2pt]
        \hline
        $f_\mathrm{ic}$         &0.92$_{-0.18}^{+0.08}$     &0.19$_{-0.07}^{+0.14}$\\
        $f_\mathrm{sat}$        &0.110$_{-0.012}^{+0.016}$  &0.151$_{-0.041}^{+0.048}$\\
        $\log_{10}\overline{M}_\mathrm{h}$&13.29$_{-0.02}^{+0.02}$&13.00$_{-0.03}^{+0.03}$\\
        $b_\mathrm{lin}$        &$2.31_{-0.04}^{+0.04}$
        &$2.13_{-0.05}^{+0.05}$\\[2pt]
        \hline        
        $\overline{n_z}$        &4.56                       &1.84\\
        \hline
    \end{tabular}
    }
    %}
    \caption{The results for the fits to high-z LRG sample with two redshift bin: $0.8<z<0.95$ and $0.95<z<1.1$. We show the mean$\pm 1 \sigma$ error for HOD and derived parameters. We also list the average comoving number density in units of $10^{-4}\ (h^{-1}\mathrm{Mpc})^{-3}$. Masses are in units of $h^{-1}M_\odot$.
      }
    \label{tab:lrgz0811}
\end{table}

\begin{figure*}
    \centering
    \includegraphics[width=0.9\linewidth]{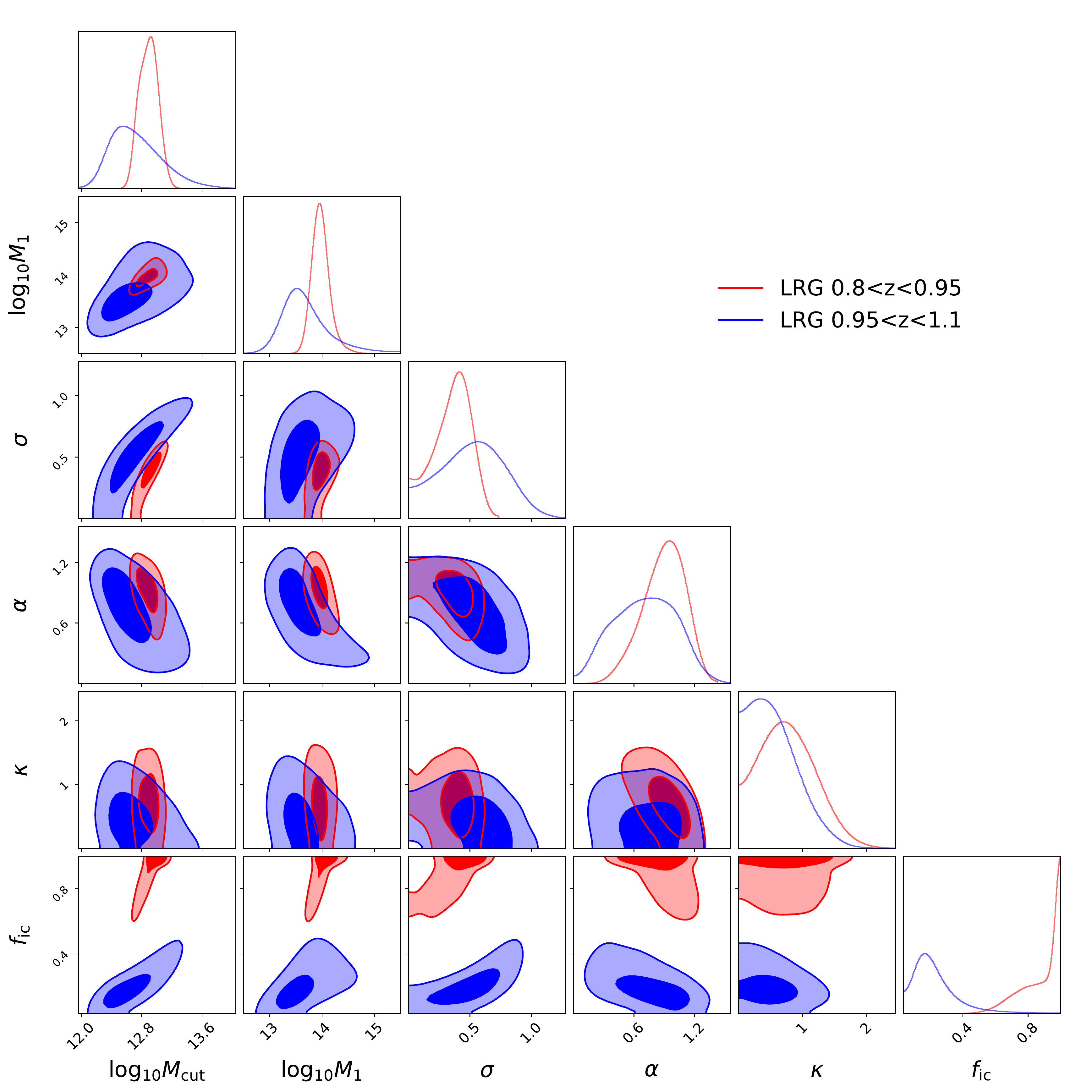}
    \caption{The DESI One-Percent Survey LRG at $z>0.8$ HOD posterior. The results from $0.8<z<0.95$ and $0.95<z<1.1$ are shown in red and blue respectively. The contours represent 68 and 95$\%$ confidence levels. 1D marginalised distribution for each parameter is shown at the top of each column.}
    \label{fig:trihodz}
\end{figure*}

\begin{figure}
    \centering
    \includegraphics[width=0.48\textwidth]{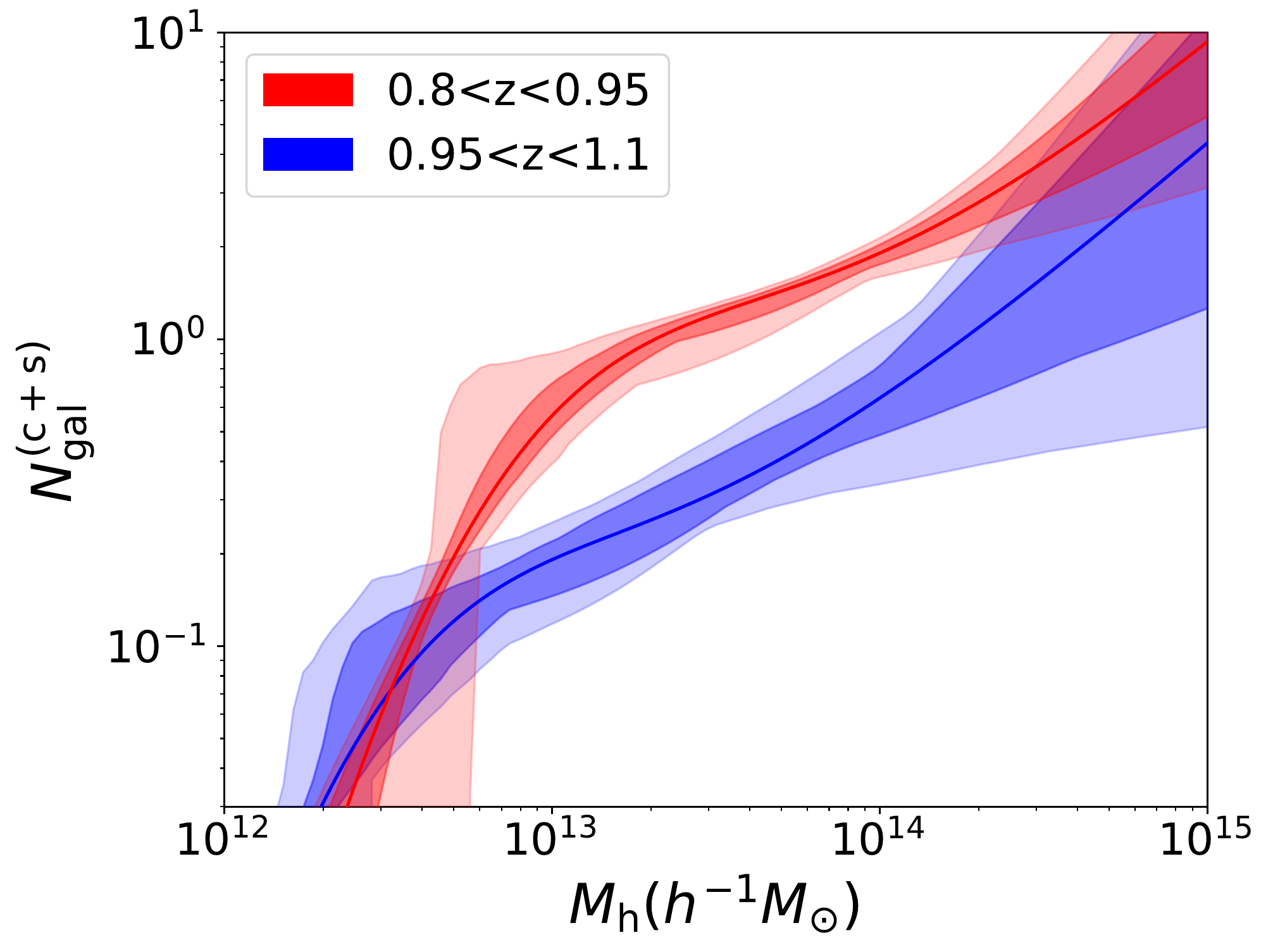}
    \caption{\protect\footnotesize The HOD posterior band (central+satellite) of LRG sample at $z>0.8$. The results from $0.8<z<0.95$ and $0.95<z<1.1$ are shown in red and blue respectively. The shaded regions correspond to 1 and 2 $\sigma$ posteriors.}
    \label{fig:bandhodz}
\end{figure}

As shown in Figure~\ref{fig:nz}, the number density of the LRG main sample experiences a significant decline beyond a redshift of 0.8, indicating a marked change in the astrophysical properties of this population. To further investigate the HOD and shed light on the evolution of the LRG sample in this redshift range, we subdivide the high-redshift LRGs into two redshift intervals, $0.8 < z < 0.95$ and $0.95 < z < 1.1$.  We use the $z=0.8$ and $z=1.1$ snapshots and employ the Zheng07+$f_\mathrm{ic}$ model to fit $w_\mathrm{p}+n_z$ data vector using the tabulation method for this aspect of the analysis.

Figure~\ref{fig:trihodz} shows the 1 and 2$\sigma$ confidence level contours for the HOD parameters. Due to the higher number density, the fit from the lower redshift interval displays a much tighter constraint as compared to the higher redshift interval, as anticipated. We find similar degeneracies between $\log M_\mathrm{cut}$ and $\sigma$, $\log M_\mathrm{cut}$ and $\log M_1$ as LRGs at $z<0.8$. The 1D distribution depicts a change in the mean value of each parameter in the Zheng07 model as the redshift increases. This trend is consistent with the comparison of the $0.4 < z < 0.6$ and $0.6 < z < 0.8$ bins in Figure~\ref{fig:corner_lrg}. The most notable difference is observed in the incompleteness parameter $f_\mathrm{ic}$. As $f_\mathrm{ic}$ can effectively control the number density of the mock galaxies, it decreases significantly as the number density of the sample decreases, indicating a drastic increase in incompleteness.

The results of our analysis provide a compelling reason for conducting a detailed study of the HOD of high-redshift LRGs. As shown in Figure~\ref{fig:bandhodz}, the 1 and 2$\sigma$ uncertainty bands of the HOD function for the high-redshift LRG sample have a minimal overlap for the range $12.8<\log M_\mathrm{halo}<14.2$, indicating a significant difference in the HOD between the two redshift intervals. Furthermore, Table~\ref{tab:lrgz0811} summarises the marginalised statistics for the high-redshift LRG main sample, which show differences in the mean completeness $f_\mathrm{ic}$ (from $92\%$ to $19\%$), the satellite fraction $f_\mathrm{sat}$ (from $11.0\%$ to $15.1\%$), mean halo mass $\log_{10}\overline{M}_\mathrm{h}$ (from $13.29$ to $13.00$) and linear bias $b_\mathrm{lin}$ (from $2.31$ to $2.13$). These results indicate that the HOD of the high-redshift LRG main sample evolves with redshift, and that DESI LRGs at $z > 0.95$ might have be a physically different sample than the lower redshift LRG sample. We describe several possible explanations in the following subsection. 

% a good place to discuss the high z sample 

\subsection{Redshift evolution of DESI LRG HOD}
\label{subsec:hodz}
\begin{figure*}
    \centering
    \includegraphics[width=0.9\linewidth]{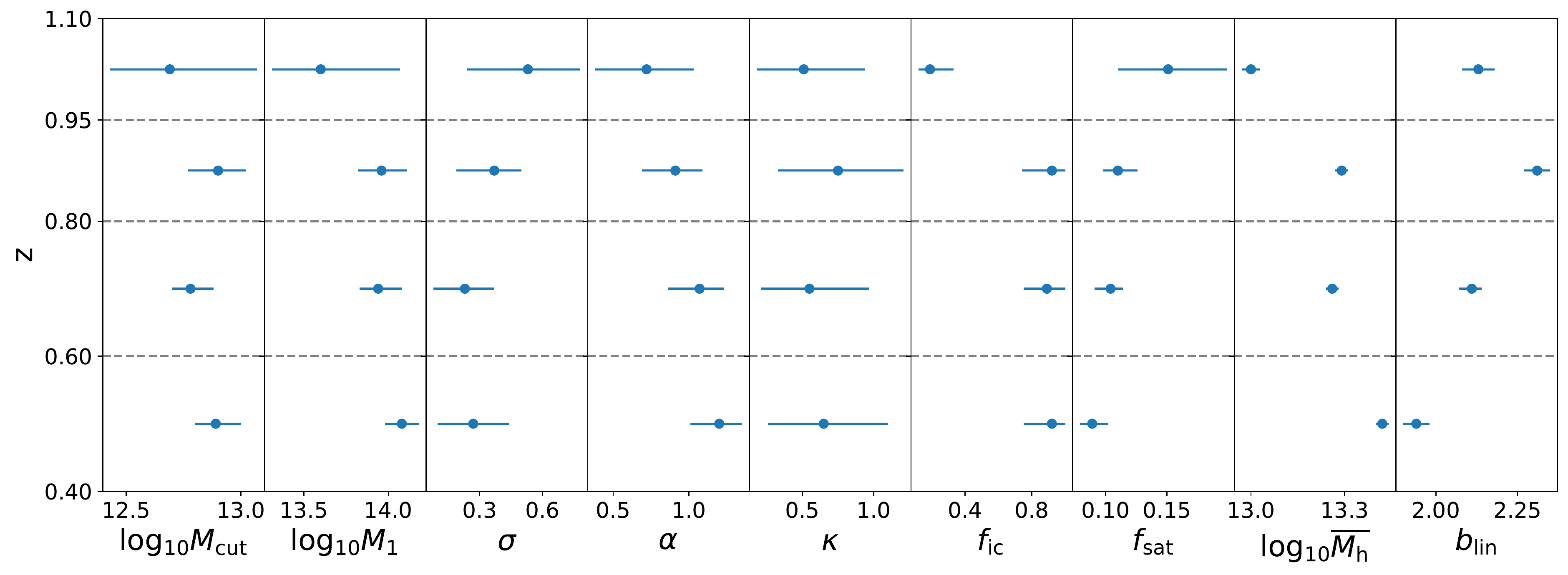}
    \caption{\protect\footnotesize The marginalised results of HOD and derived parameters of LRG main sample for each redshift bin. Markers and error bars show the mean of the fits and 1$\sigma$ error. The numerical results of this plot are also listed in Table~\ref{tab:LRG_fit1} and \ref{tab:lrgz0811}.}
    \label{fig:boxz}
\end{figure*}

The evolution of HOD and derived parameters of LRGs across all redshift bins are shown in Fig.~\ref{fig:boxz}. We only use data vector $w_\mathrm{p}+n(z)$ in this part of the analysis for consistency. The central galaxy host halo mass threshold ($M_\mathrm{cut}$) does not exhibit any significant trend with redshift within the uncertainties. The scatter in the halo mass threshold ($\sigma$) tends to increase with increasing redshift, indicating a smoother transition in the central HOD on high redshift. In terms of satellite parameters, both $M_1$ and $\alpha$ show a declining trend with redshift, while $\kappa$ shows little variation with redshift and is weakly constrained by the data.

The satellite fraction ($f_\mathrm{sat}$) shows a mild trend of increasing with redshift at $z < 0.95$, rising from $f_\mathrm{sat}\simeq 9\%$ at $z\simeq0.5$ to $f_\mathrm{sat}\simeq 11\%$ at $z\simeq 0.9$. This increase in satellite fraction can be interpreted as a result of the merging of galaxies over time. At $z > 0.95$, the satellite fraction increases substantially to $15\%$, but the significance is low. 

The mean halo mass does not exhibit a clear trend except at $z > 0.95$, where the mean halo mass drops off substantially. A similar behavior is seen for the galaxy bias, which increases with redshift at $z < 0.95$ (consistent with studies of DESI-like photometric LRGs in \cite{2020Zhou}) but drops off at $z > 0.95$. Furthermore, the highest redshift bin also exhibits a drastic drop in completeness. These comparisons serve as strong evidence that that the LRG sample at $z > 0.95$ is physically different from the LRGs at $z < 0.95$. 

\cite{2022Zhou} showed that the DESI LRG selection should yield a highly stellar mass complete sample at $\log_{10}M_* > 11.5$ between redshift $0.4 < z < 0.9$. At $z > 0.9$ the completeness at the high mass end starts to deviate from 1 and drops further at $z > 1$. This point is confirmed by explicitly computing the stellar mass functions of DESI SV3 LRGs via SED fitting in \textcolor{blue}{Gao et al. in prep}, where a substantial incompleteness at $\log_{10}M_* > 11.5$ is observed at $z > 1$. This incompleteness at the high mass end can partially explain the drop in the inferred bias and halo mass in the highest redshift bin. 

Another potential contributing effect is that the DESI LRG selection employs a sliding colour-magnitude cut in $r-W1$ vs $W1$ that turns over at $W1=\sim -19$ to include more galaxies at the high redshift end (see Figure~3 of \cite{2022Zhou}). However, this turn over also includes red galaxies with very faint $W1$ magnitudes into the LRG sample, thus possibly decreasing the mean halo mass and mean galaxy bias. 

Separately, \cite{2023Setton} inferred star formation histories from DESI SV1 LRG SEDs and found evidence that recently quenched (post-starburst) galaxies constitute a growing fraction of the massive galaxy population with increasing redshift. The study showed that these galaxies are significantly brighter than the parent LRG sample at fixed stellar mass. Thus, at fixed brightness, these galaxies likely have lower halo masses and lower biases compared to the parent LRG sample. If these galaxies are indeed a significant fraction of high redshift LRGs, then they would contribute to the drop in the mean halo mass and bias. This could also explain the relative high satellite fraction as recently accreted satellites are also likely to have been recently quenched. 

% Additionally, the selection process of LRGs at high redshift tends to favor bluer galaxies, which signifies that the selected LRG samples might not be as representative of galaxies as massive as those at lower redshifts, consequently decreasing galaxy bias.
%The galaxy bias progressively escalates with an increase in redshift, increasing from $b\simeq1.9$ at $z\simeq0.5$ to $b\simeq2.3$ at $z\simeq0.9$. Nevertheless, we observe a decline in galaxy bias for the highest redshift bin, with $b\simeq2.1$ at $z\simeq1.0$.

%However, we note that one important caveat in the interpretation of the mean halo masses is the use of fixed redshift snapshots in this analysis. 
An important caveat to consider in the interpretation of the mean halo masses relates to the use of fixed redshift snapshots in this analysis.
%Although snapshots-based \textsc{AbacusHOD} can generate high-fidelity mocks that accurately recover the clustering measurements of the data, it can be problematic for understanding the redshift evolution of HOD. 
To obtain an unbiased HOD parameter estimation, the mean redshift of the redshift bin must remain close to the snapshot redshift. However, the choice of snapshots is limited to several primary redshifts provided by the \textsc{AbacusSummit} simulation. For instance, we use the snapshot at $z=0.8$ for both $0.6<z<0.8$ and $0.8<z<0.95$ redshift bins, resulting in mean sample redshifts that are lower and higher than the snapshot redshift, respectively. 
%When the mean redshift is lower than the snapshot redshift, the simulation lacks the necessary clustering power to match the data, leading to an overestimation of parameters such as $\log M_\mathrm{cut}$ and $\log \overline{M}_\mathrm{h}$. Conversely, when the mean redshift is higher than the snapshot redshift, the simulation has more clustering power than the data, leading to an underestimation of those parameters. This effect is particularly pronounced for the $0.95<z<1.1$ redshift bin, as the drop in number density results in a more significant bias of the mean sample redshift from the snapshot redshift of $z=1.1$.
As a result, the mean halo mass has been overestimated for $0.8<z<0.95$ and underestimated for $0.6<z<0.8$. This effect is particularly pronounced for the $0.95<z<1.1$ redshift bin, as the drop in number density results in a more significant bias of the mean sample redshift from the snapshot redshift of $z=1.1$. Mitigating this effect could, to some extent, resolve the non-monotonic shape in the mean halo mass evolution. 
Nevertheless, we maintain that this effect does not fully explains the drop in halo mass in the highest redshift bin because the drop in linear bias is agnostic to simulation snapshots as the underlying clustering amplitude is computed at the mean redshift of the sample instead of the simulation snapshot. 

%However, %
This systematic highlights the need for future redshift evolution studies to use more accurate redshifts. The shift in the mean value of HOD and derived parameters also provide a strong scientific incentive to conduct a detailed study of LRGs using the halo light-cone catalogs in combination with a redshift-evolved HOD model.

% \subsection{ELG}

% If we just fit wp with a HMQ model, fixing Q, and a power law satellite occupation, then the best fit fails to capture the small scales. If we relax the satellite occupation with a roll off at lower masses, the resulting chi2 remains the same. So just judging from wp, the satellite roll off is not preferred. wp also does not favor the inclusion of a central error piece. 

\section{QSO HOD results}
\label{sec:qso}
We analyze the QSO sample following the same procedure. First, we construct the mock-based covariance matrix by fitting a baseline HOD on the QSO \xirppi measurement. However, we find that directly fitting \xirppi returns poor fits because the QSO \xirppi measurement below $r_p\sim 1\ $\hmpc\ have particularly low signal-to-noise, and the corresponding jackknife errors do not behave properly. 
%\sandy{This is probably something we need to elaborate on and understand more. Ashley said this might be due to differences between gaussian and poisson. I dont quite understand?} 
Instead, we fit just the projected $w_p$ with the 5-parameter+incompleteness model, where we do achieve a good fit. Then we independently tune the two velocity bias parameters to match the \xirppi signal at $r_p > 1\ $\hmpc. We again obtain a good fit. With that, we populate the 1800 small boxes and generate a mock-based covariance matrix for QSO \xirppi.

% Directly fitting the 2D xirppi turns out to return very poor fits. This is because the xirppi measurement below 1mpc/h has very poor signal-to-noise, except at the lowest rpi bin. So we first run optimization of a base 5-parameter HOD on the projected wp. Then we fit the xirppi measurement at above 1mpc rp, but only varying the velocity bias parameters. We take this best fit to compute covariance matrix. 

\begin{figure*}
    \hspace{-0.7cm}
    \includegraphics[width=0.9\textwidth]{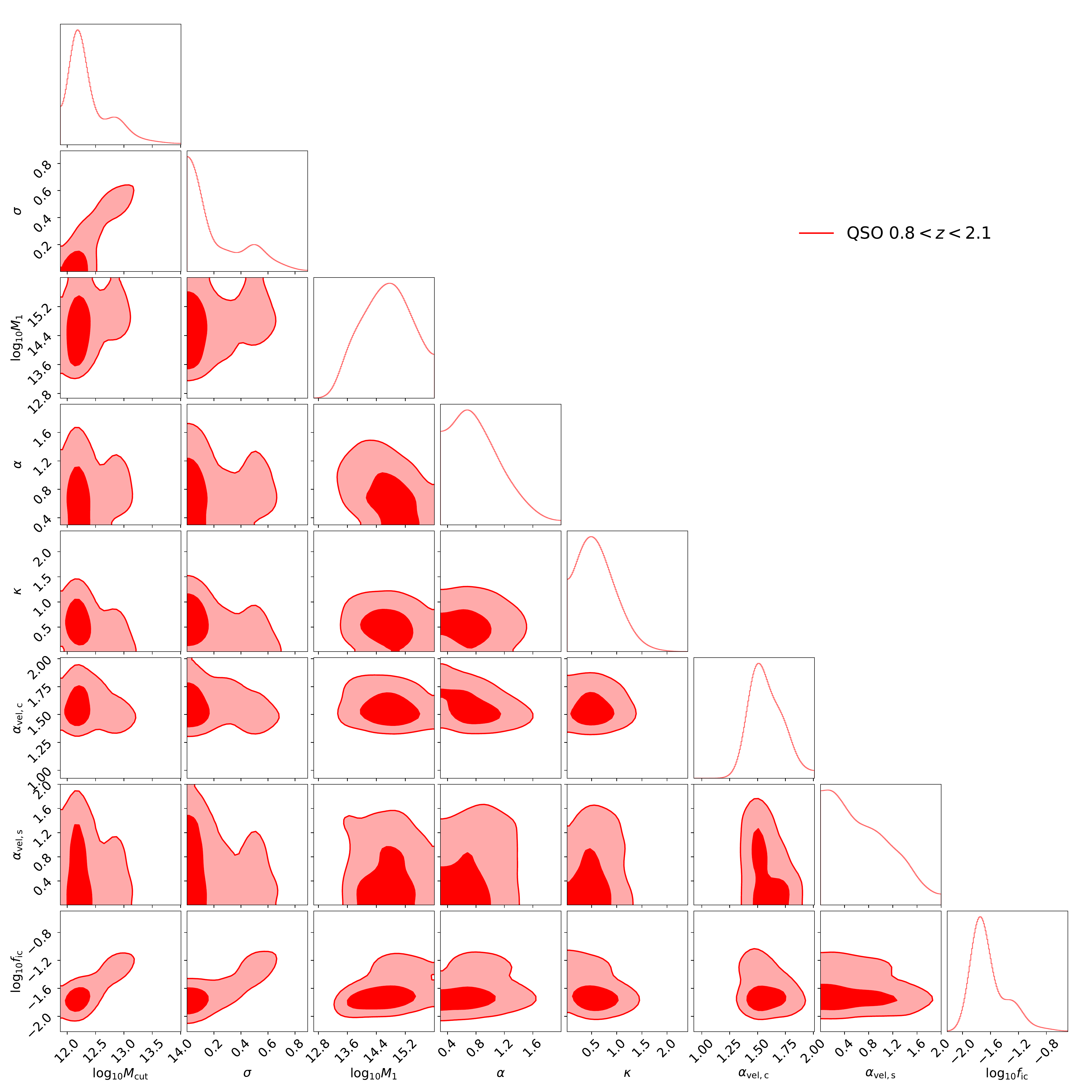}
    \vspace{-0.2cm}
    \caption{The DESI One-Percent Survey QSO HOD posterior. The red contours correspond to the 1 and 2$\sigma$ marginalised posteriors for QSOs in the $0.8 < z < 2.1$ bin.}
    \label{fig:corner_qso}
\end{figure*}

\begin{figure*}
         \hspace{-0.7cm}
         \centering
         \includegraphics[width=\textwidth]{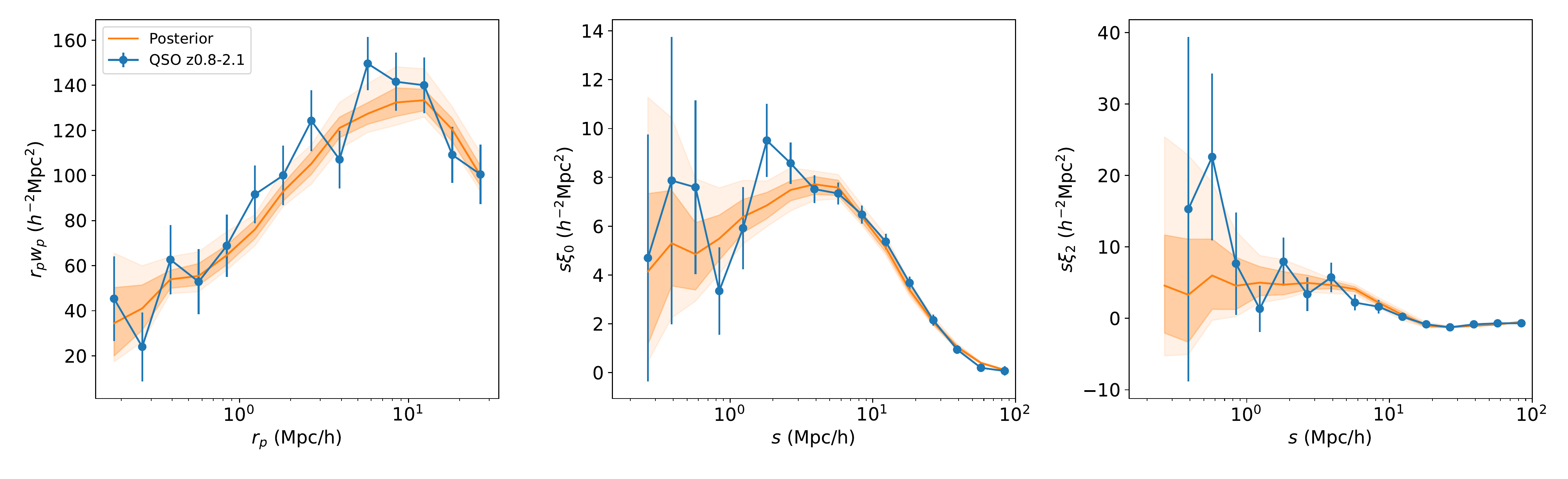}
        \vspace{-0.4cm}
    \caption{The QSO $w_p$ and multipoles posterior predictives compared to the data. The blue lines correspond to the One-Percent Survey measurement with jackknife error bars. The solid orange line denote the prediction corresponding to the posterior mean. The orange shaded regions correspond to the 1 and 2$\sigma$ full posterior using the full mock covariance matrix.}
    \label{fig:multipoles_qso}
\end{figure*}

\begin{figure}
         \hspace{-0.4cm}
         \centering
         \includegraphics[width=0.48\textwidth]{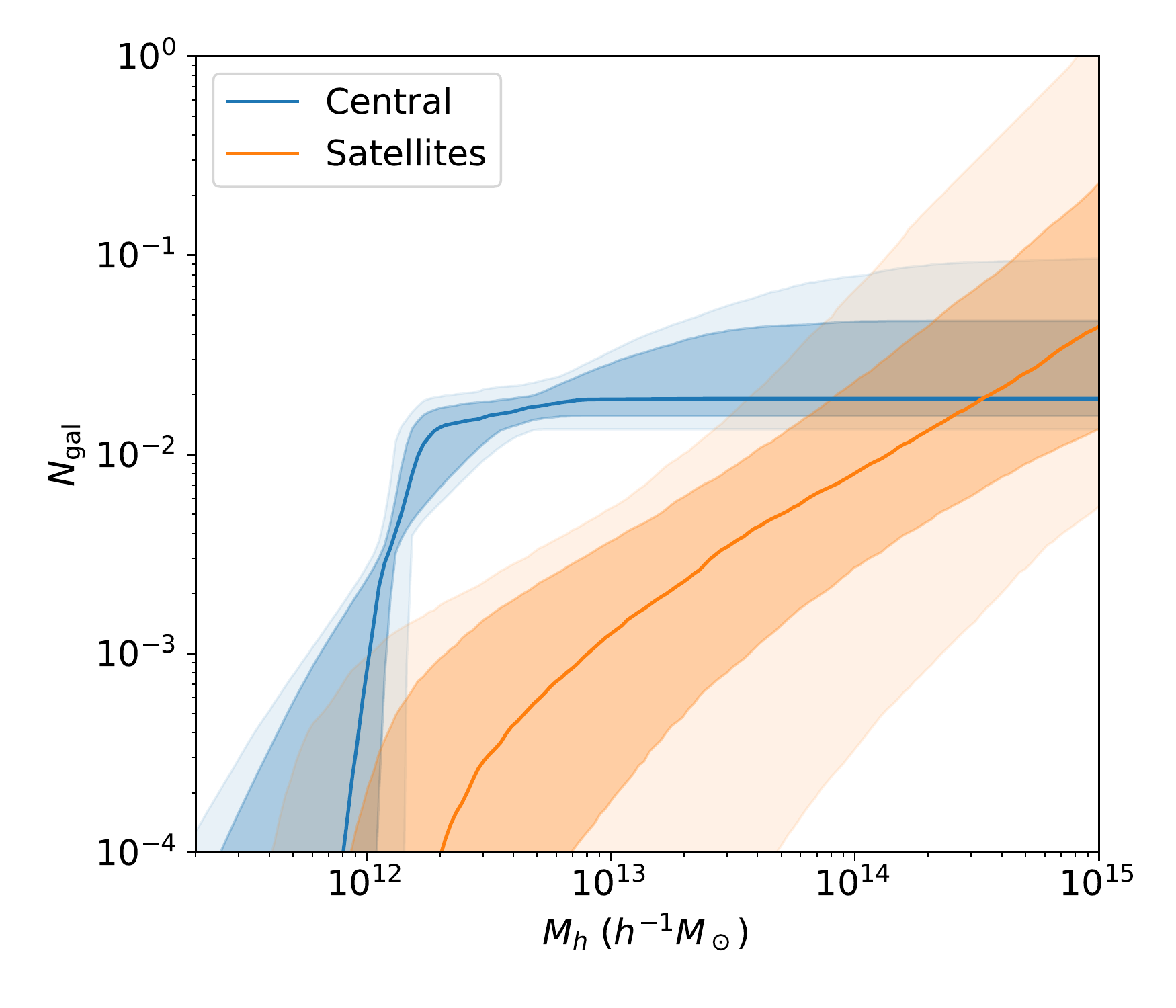}
        \vspace{-0.4cm}

    \caption{The HOD posterior for the QSO sample. The shaded regions correspond to 1 and 2$\sigma$ posteriors. }
    \label{fig:hod_qso}
\end{figure}

Again, we sample the baseline HOD parameter to obtain marginalised posteriors for the QSO HOD. The results are visualised in Figure~\ref{fig:corner_qso} and summarised towards the bottom of Table~\ref{tab:LRG_fit1}. Figure~\ref{fig:hod_qso} shows the corresponding HOD posterior. In general, the QSO HOD parameters are much less constrained than the LRG parameters due to the limited sample size. For the same reason, we also do not test additional model extensions as we already achieve an excellent best-fit $\chi^2$ with the baseline model. We will conduct such tests when a significantly larger sample of QSOs become available. We also showcase the $w_p$-only constraints in Table~\ref{tab:LRG_fit1}, and we find them to be consistent with the \xirppi constraints. 

The HOD constraints compare well with those inferred for eBOSS QSOs as presented in Table~1 of \cite{2020Alam}. Specifically, they found $\log M_\mathrm{cut} = 12.2$ and $\log M_1 = 14.1$, consistent with our findings. Perhaps the most unexpected parameter constraint is the central velocity bias, where we find \smash{$\alpha_c = 1.54^{+0.17}_{-0.08}$}, meaning that the central galaxies exhibit large peculiar velocities relative to the halo center. There are several potential explanations for this. While one possibility is that it points towards energetic processes within the AGN, we speculate that this may also be due to the rather large redshift uncertainties with the QSO sample (\textcolor{blue}{DESI Collaboration in prep}). Specifically, QSO primary spectral lines such as Mg~\textsc{ii} line to C~\textsc{iv} suffer from large systematic velocity shifts caused by astrophysical effects \citep[e.g.,][]{2018Zarrouk, 2011Richards, 2002Richards}. Another potential explanation is that QSOs preferentially occupy recently merged halos, resulting in large velocity dispersion relative to the halo center-of-mass. In the following paragraph, we also show that uncertainties in satellite fraction can also be degenerate with velocity bias. Thus, the large velocity bias in the QSO sample could be due to redshift uncertainties, AGN physics and mergers, and/or uncertainties in satellite fraction. 

In terms of derived quantities, we infer a mean halo mass of \smash{$\log_{10}\overline{M}_h= 12.65^{+0.09}_{-0.04}$} for the DESI QSO sample, consistent with the $\log_{10}\overline{M}_h= 12.7$ found for the eBOSS QSO sample in both \cite{2017Rodriguez-Torres} and \cite{2020Alam}. We report a linear bias of \smash{$b_\mathrm{lin}=2.63^{+0.37}_{-0.26}$} at $z\simeq1.5$ for the DESI QSO sample, which is slightly higher than the first-year eBOSS quasar sample with \smash{$b_\mathrm{lin}=2.45\pm0.05$} at $z=1.55$, as found by \citet{2017Laurent}. However, considering the error bar, the results are consistent. We infer a satellite fraction of \smash{$3^{+8}_{-2}\%$} for the QSO sample, consistent with $f_\mathrm{sat} = 5\%$ found in \cite{2017Rodriguez-Torres} using a subhalo abundance matching model but significantly lower than the $30\%$ inferred with a multi-tracer HOD fit in \cite{2020Alam}. The large satellite fraction inferred in \cite{2020Alam} is directly a result of a small inferred $\alpha = 0.4$, which results in a large number of satellites in low mass halos. However, we speculate that this result might in fact be degenerate with our finding as most of these satellites in low mass halos do not have companion centrals due to the low completeness. These ``rogue'' satellites are ill defined in the vanilla HOD context, and can simply be re-classified as centrals. This could also be connected to the large central velocity bias we found, which could disappear if we re-classify some centrals as satellites, which would naturally have larger velocity dispersion. Our results compare well with earlier SDSS Quasar HOD fits. \cite{2013Shen} obtained a satellite fraction of 7-10$\%$ and found the inferred satellite fraction to be dependent on the assumed HOD model. \cite{2012Richardson} obtained a lower satellite fraction of $\sim 0.1\%$ for $z\sim 1.4$ quasars.

These types of questions also show that the QSO--halo connection physics is poorly understood. In fact, we use the standard 5-parameter model in this analysis simply because we have no evidence that a different model is favored. We reserve a more comprehensive analysis of QSO HOD for a future paper when a much larger sample of QSOs becomes available. 

% show HOD plots

\section{Mock products}
\label{sec:mocks}
% cubic mocks
We apply the best-fit HODs obtained for the LRG and QSO samples to all 25 base boxes available in \textsc{AbacusSummit} at Planck cosmology to create high fidelity mocks. For each tracer at each redshift snapshot, the total sample volume is 200$h^{-3}$Gpc$^{3}$ comoving. The volume provided by \textsc{AbacusSummit} is an order of magnitude larger than other simulations of comparable resolution, reaching 5-10 times the volume expected to be observed by DESI. Given the volume and resolution, these mocks are critical for testing and calibrating DESI cosmology pipelines at the necessary precision. 

% variation mocks
In addition to using just the best-fit HODs, we also create additional mocks where we perturb the HOD parameters around the best-fit to generate mocks that share the same cosmology but differ in bias prescriptions. The perturbations are sampled from the $3\sigma$ region of the parameter space around the best-fit values. We repeat this procedure for both the baseline HOD model and an extended HOD model that also includes environment-based assembly bias $B$ and satellite radial profile parameter $s$, resulting in a set of mocks that encompass a diverse range of possible HODs. 

These variety mocks enable key robustness tests of large-scale cosmology inference pipelines against galaxy--halo connection systematics. Specifically, large-scale cosmology pipelines that utilise the BAO/RSD features or full-shape information often assume a much simpler bias model. As a result, complexities in galaxy--halo connection modeling could become degenerate with cosmology and thus result in systematic bias in the inferred cosmology. Thus, it is essential to test cosmology inference pipelines against a range of mocks with varying bias models and demonstrate that the cosmology inference remains unbiased. We defer a detailed discussion of these tests to a dedicated paper (\textcolor{blue}{DESI Collaboration in prep}).

% 1.27203e+01 1.371603e+01 -5.995368e-01 1.3227e+00 4.7554e-01 2.10234e-01 8.34229e-01 0.0000000e+00 0.00000+00 0.0000000e+00
% 1.26861e+01 1.371867e+01 -6.415894e-01 1.3094e+00 3.2901e-01 1.48986e-01 9.30799e-01 0.0000000e+00 0.00000+00 0.0000000e+00
% 1.27690e+01 1.371758e+01 -5.243963e-01 1.1835e+00 5.2292e-01 1.18379e-01 9.32456e-01 0.0000000e+00 0.00000+00 0.0000000e+00
% 1.26594e+01 1.392126e+01 -2.229869e+00 1.2968e+00 2.4551e-01 1.28932e-01 9.04570e-01 1.5033013e-01 -9.7055e-01 4.7192100e-02
% 1.26254e+01 1.385800e+01 -1.703135e+00 1.1939e+00 7.9151e-01 2.10991e-01 9.10010e-01 7.9179725e-02 -6.5193e-01 -2.30626e-02
% 1.26369e+01 1.391530e+01 -1.651592e+00 1.1034e+00 8.4892e-01 1.52895e-01 7.91337e-01 1.2105774e-01 -8.3441e-01 -3.59013e-01

% lightcone mocks
In addition to constructing cubic mocks, we utilise our best fits to construct redshift-dependent mocks on the \textsc{AbacusSummit} lightcones \citep{2022Hadzhiyska}. The benefit of having mocks on the lightcone is that they provide an accurate synthetic map of the sky, which is crucial for testing out systematic and observational effects such as fiber collisions. Lightcones also enable explicit modeling of redshift evolution, such as in the forward modeling pipelines being developed for novel summary statistics \citep[e.g.,][]{2023Yuan, 2022HahnForward}. Our procedure for generating lightcone mocks is as follows: 
\begin{enumerate}
    \item Adopting the \ahod\ algorithm, we first subsample the halo lightcone catalogues assuming the same envelope as the cubic boxes and pre-compute various assembly bias parameters and decorations to the HOD model.
    \item We read in the best-fit parameters as we found in section~\ref{sec:results}. We linearly interpolate the HOD parameters as a function of redshift, pivoting on the two best-fits in the two redshift bins. 
    \item Finally, we generate the galaxy mocks for all tracers/the LRGs at all available redshifts of the 25 fiducial cosmologies, i.e. \texttt{AbacusSummit\_c000\_ph000-024}. We note that due to the geometry, each of our mocks provides an octant of the sky until $z \approx 0.8$, decreasing gradually as we go to higher redshifts.
\end{enumerate}

All said mocks will be made publicly available at a future date as a part of DESI EDR. 
%\section{Discussions}
%\label{sec:discuss}

%In this section, we introduce the tabulation method for AbacusSummit-based HOD analysis, which has been previously proposed to significantly reduce the computational time for HOD fitting. Then, we analyse and present the HOD of the high-redshift LRG sample ($z>0.8$), where the number density drop significantly. Finally, we discuss the evolution of the LRG main sample across redshifts. 

\section{Conclusions}
\label{sec:conclude}
In this paper, we present a comprehensive analysis of the halo occupation distribution of the DESI One-Percent Survey LRG and QSO samples using the \textsc{AbacusSummit} cubic boxes. 

For LRGs, we study the sample in two fiducial redshift bins $0.4 < z< 0.6$, $0.6 < z< 0.8$, and also in a third high redshift bin $0.8 < z < 1.1$. In the fiducial bins, we compare the baseline HOD model with extended models with galaxy assembly bias and satellite profile bias. We find no evidence for model extensions at current precisions and the baseline model is favored by the data. For both redshift bins, we constrain the baseline parameter posteriors with the \xirppi data vector and mock-based covariance matrix. The resulting model posteriors produce the correct redshift-space clustering. We find strong constraints on inferred properties such as the average halo mass and the satellite fraction, which are broadly consistent with results from eBOSS and BOSS. We also find consistency between \xirppi fit and $w_p$-only fit. The marginalised posterior constraints are summarised in Table~\ref{tab:LRG_fit1}.
To highlight a few key constraints: the LRG sample in $0.4 < z < 0.6$ yields a satellite fraction of $11\pm 1\%$ and a mean halo mass of \smash{$\log_{10}\overline{M}_h=13.40^{+0.02}_{-0.02}$}, whereas the $0.6 < z < 0.8$ sample results in a satellite fraction of $14\pm 1\%$ and a mean halo mass of \smash{$\log_{10}\overline{M}_h=13.24^{+0.02}_{-0.02}$} in $0.6 < z < 0.8$.

Combining the fiducial analysis at $z < 0.8$ and high redshift analysis at $z>0.8$, we find clear trends of evolution, especially in physical parameters like satellite fraction and mean halo mass. Specifically the mean halo mass decreases with redshift whereas the satellite fraction increases with redshift. This motivates future redshift evolution studies that should shed light on the physics of the evolution of massive galaxies. The marginalised posterior constraints are summarised in Table~\ref{tab:lrgz0811}. We also find the properties of the LRG sample at $z > 0.95$ to deviate significantly from lower redshift LRG sample. Specifically, the LRGs at $z > 0.95$ display significantly lower mean halo mass, a very low completeness, a lower mean bias while showing a somewhat high satellite fraction. We offer a few plausible explanations of these differences. 

The QSO sample is limited by sample size, and we are not able to conduct meaningful comparisons between different HOD models. Regardless, we derive good fits on the data and present posterior constraints. The marginalised posterior constraints are summarised in Table~\ref{tab:LRG_fit1}. We infer a satellite fraction of \smash{$3^{+8}_{-2}\%$} and a mean halo mass of \smash{$\log_{10}\overline{M}_h= 12.65^{+0.09}_{-0.04}$} in redshift range $0.8 < z < 2.1$. The inferred mean halo mass is consistent with previous results, but there is some discrepancy in the satellite fraction. We speculate that such discrepancy is model dependent and we intend to revisit of this issue when a significantly larger sample of QSOs become available. 

Finally, we leverage our HOD fits to generate a large suite of DESI-like mocks. We highlight mocks with varied HODs that test the robustness of large-scale cosmology pipelines, and lightcone-based mocks that are important for building realism and testing observational systematics.

% \sout{Finally, we introduce the tabulation scheme that accelerates HOD calculation. We utilize this scheme to analyze the redshift evolution of the LRG sample at $z > 0.8$.}  

\section*{Acknowledgements}

This material is based upon work supported by the U.S. Department of Energy (DOE), Office of Science, Office of High-Energy Physics, under Contract No. DE–AC02–05CH11231, and by the National Energy Research Scientific Computing Center, a DOE Office of Science User Facility under the same contract. Additional support for DESI was provided by the U.S. National Science Foundation (NSF), Division of Astronomical Sciences under Contract No. AST-0950945 to the NSF’s National Optical-Infrared Astronomy Research Laboratory; the Science and Technologies Facilities Council of the United Kingdom; the Gordon and Betty Moore Foundation; the Heising-Simons Foundation; the French Alternative Energies and Atomic Energy Commission (CEA); the National Council of Science and Technology of Mexico (CONACYT); the Ministry of Science and Innovation of Spain (MICINN), and by the DESI Member Institutions: \url{https://www.desi.lbl.gov/collaborating-institutions}. Any opinions, findings, and conclusions or recommendations expressed in this material are those of the author(s) and do not necessarily reflect the views of the U. S. National Science Foundation, the U. S. Department of Energy, or any of the listed funding agencies.

The authors are honored to be permitted to conduct scientific research on Iolkam Du’ag (Kitt Peak), a mountain with particular significance to the Tohono O’odham Nation.

%%%%%%%%%%%%%%%%%%%%%%%%%%%%%%%%%%%%%%%%%%%%%%%%%%
\section*{Data Availability}

 The simulation data are available at \url{https://abacussummit.readthedocs.io/en/latest/}. The \ahod\ code package is publicly available as a part of the \textsc{abacusutils} package at \url{http://https://github.com/abacusorg/abacusutils}. Example usage can be found at \url{https://abacusutils.readthedocs.io/en/latest/hod.html}.
All mock products will be made available at \url{https://data.desi.lbl.gov}.

The MCMC chains generated, along with the clustering measurements used in this study - encompassing correlation functions and covariance matrices - are available in a machine-readable format at \url{https://doi.org/10.5281/zenodo.7972386}.



\bibliographystyle{mnras}
\bibliography{biblio} % if your bibtex file is called example.bib

%\begin{thebibliography}{99}
%\end{thebibliography}

%%%%%%%%%%%%%%%%%%%%%%%%%%%%%%%%%%%%%%%%%%%%%%%%%%

%%%%%%%%%%%%%%%%% APPENDICES %%%%%%%%%%%%%%%%%%%%%

\appendix

\section{Accelerating HOD fitting with tabulation method}
\label{app:tab}

In the standard procedure for fitting the HOD model, galaxy mock populations are generated for each set of HOD parameters that have been sampled. The clustering statistic of interest is subsequently measured and compared to measurements obtained from the data in order to estimate the likelihood. This procedure is repeated multiple times for various HOD parameters until the posterior likelihood has been thoroughly explored. An alternative approach, known as the tabulation method, was first proposed by \citet{2011MNRAS.416.1486N} and later expanded upon by \citet{2016MNRAS.458.4015Z}. This method reverses the order of applying the HOD model and measuring the clustering. Specifically, the clustering of halos is precomputed prior to the Markov Chain Monte Carlo (MCMC) stage, and the HOD population scheme is subsequently applied by combining weights with the halo clustering. This approach significantly improves the efficiency of the HOD fitting process, as the most computationally intensive step is moved outside of the MCMC loop. We summarise the tabulation method for the \textsc{AbacusSummit} simulation below but refer the reader to \citet{2022MNRAS.515.6133Z}. for more details.

We take projected 2PCF $w_\mathrm{p}$ here as an example and assume our HOD model only depend on the mass of the host halo. To match the behavior of \ahod, where we populate the satellite with particles, we first divide the halo catalogue and particle catalogue attached to the halos into $N_\mathrm{b}$ bins. Then the galaxy correlation function $w_{\mathrm {p,gg}}$ is given by a weighted sum over different mass bin cross-correlations.

\begin{equation} \label{eqn:wptab}
    \begin{aligned}
        w_{\mathrm {p,gg}}(r_\mathrm{p})&=\sum_{i,j}^{N_\mathrm{b}}w_\mathrm{cent}(M_i)w_\mathrm{cent}(M_j)w_{\mathrm{p,hh}}(r_\mathrm{p},M_i,M_j)\\
                &+2\sum_{i,j}^{N_\mathrm{b}} w_\mathrm{cent}(M_i)w_\mathrm{sat}(M_j)w_{\mathrm{p,hp}}(r_\mathrm{p},M_i,M_j)\\
                &+\sum_{i,j}^{N_\mathrm{b}} w_\mathrm{sat}(M_i)w_\mathrm{sat}(M_j)w_{\mathrm{p,pp}}(r_\mathrm{p},M_i,M_j),
    \end{aligned}
\end{equation}
where $w_{\mathrm{p,hh}}(r_\mathrm{p},M_i,M_j)$ is the two-point cross-correlation function of halos in the $i$th and $j$th mass bins (similarly for the halo-particle and particle-particle correlation functions) and naively we could take the weight as
\begin{align}
    w_{\mathrm{cent}}(M_i) &= \bar{n}_{\mathrm{cent}}(M_i)\label{eqn:wcen},\\
    w_{\mathrm{sat}}(M_i) &= \bar{n}_{\mathrm{sat}}(M_i)\frac{N_\mathrm{h}^i}{N_\mathrm{p}^i}\label{eqn:wsat}.
\end{align}
where $N_\mathrm{h}^i$ and $N_\mathrm{p}^i$ are numbers of halos and particles in $i$th mass bin.

Equation~\ref{eqn:wptab} gives an average value of clustering expected for a given HOD model instead of a specific realisation that includes stochastic noise and further reduces the bias of HOD fitting. We also test and confirm that the mean prediction is consistent with the output of the standard \ahod\ approach. 

While the tabulation method can accelerate the fitting of HOD, it also limits the flexibility of extending the HOD model. The baseline HOD model relies solely on halo mass, making it easy to prepare tabulated halo and particle pair counts across mass bins. However, when dealing with more complex HOD models that involve velocity bias and assembly bias extensions, additional dimensions of dependency can significantly increase the complexity of preparing tabulated halo and particle pair counts. Therefore, we only employ the tabulation method to quickly evaluate the HOD posterior when using the Zheng07+$f_\mathrm{ic}$ model to fit the $w_\mathrm{p}+n_z$ data vector. For all other cases, we use the standard \textsc{AbacusHOD} approach.

\section{choice of prior}
\label{app:prior}

\begin{figure*}
    \centering
    \includegraphics[width=0.7\linewidth]{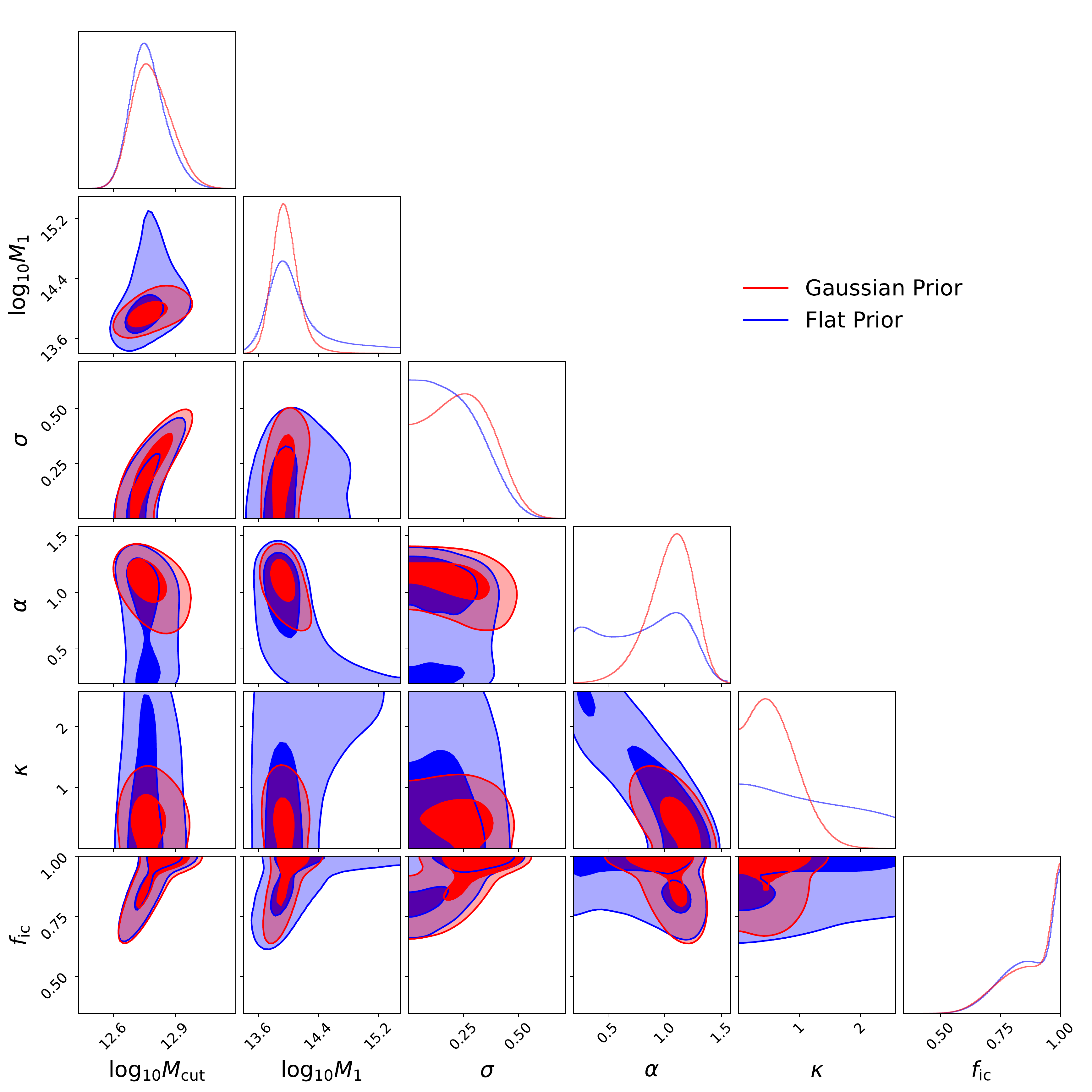}
    \caption{\protect\footnotesize Marginalised probability distribution of HOD parameters for LRG sample at $0.6<z<0.8$ with and without Gaussian priors listed in Tab.~\ref{tab:priors}. The results using Gaussian prior and flat prior are shown in red and blue respectively. The contours represent 68 and 95$\%$ confidence levels. 1D marginalised distribution for each parameter is shown at the top of each column.}
    \label{fig:triprior}
\end{figure*}

\begin{figure}
    \centering
    \includegraphics[width=0.45\textwidth]{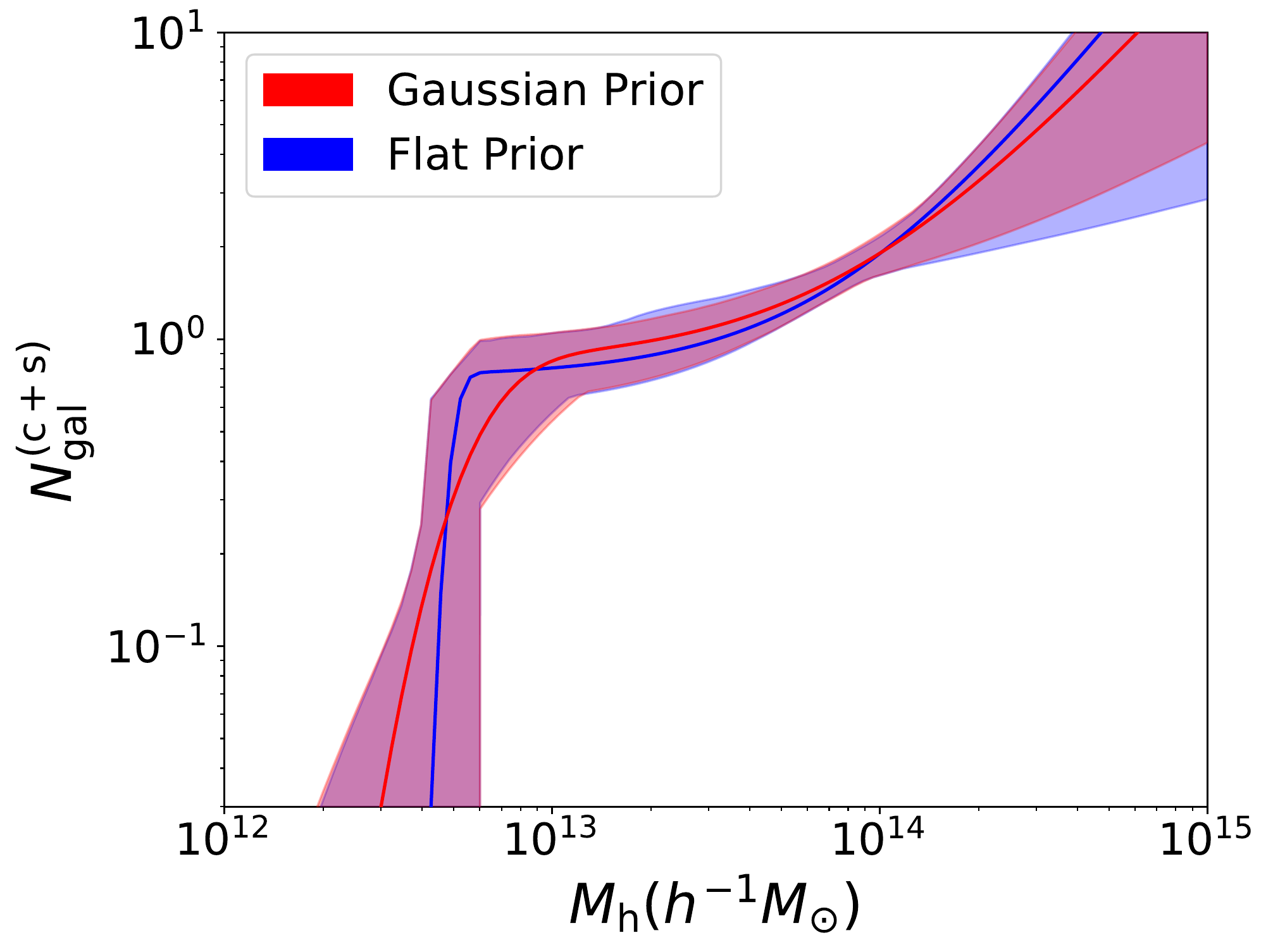}
    \caption{\protect\footnotesize 2$\sigma$ band of LRG sample HOD at $0.6<z<0.8$ with and without Gaussian prior. The red is the 95$\%$ CL uncertainty from fit using Gaussian priors, the blue is the 95$\%$ CL uncertainty from fit using flat priors. Lines are the corresponding best fit.}
    \label{fig:2sigmaprior}
\end{figure}

The selection of priors for the HOD parameters can potentially impact the inferred constraints. In order to eliminate extreme values of the HOD parameters and gain a more comprehensive understanding of the Galaxy-Halo connection model, Gaussian priors shown in Tab.~\ref{tab:priors} were applied to the HOD parameters in the primary analysis. In order to assess the potential impact of these priors on the results, additional MCMC analyses were conducted using only flat priors. 

In this test, we employ the Zheng07+$f_\mathrm{ic}$ model to fit $w_\mathrm{p}+n_z$ of LRG at redshift $0.6<z<0.8$. Figure~\ref{fig:triprior} illustrates the 1 and 2$\sigma$ confidence level contours for the HOD parameters when different priors are applied. It is clear that the fit with a Gaussian prior displays a tighter contour, particularly for the satellite parameters $\log_{10} M_1$, $\alpha$, and $\kappa$. The Gaussian prior helps to eliminate isolated regions in the contour, where the data have limited power to constrain the parameters. The 1D distribution of each parameter in Figure~\ref{fig:triprior} indicates that the Gaussian prior has a smaller impact on the central parameters as compared to the satellite parameters and hardly any impact on incompleteness. 

Figure~\ref{fig:2sigmaprior} shows the 2$\sigma$ uncertainty band of HOD posterior and best fit for both cases. The two bands overlap heavily with each other. The Gaussian prior fit has a slightly narrower band for $\log_{10} M_\mathrm{cut} > 14.2$, where the satellite parameters have a larger impact on the HOD. Additionally, the Gaussian prior presents a smoother best fit than the flat prior, which exhibits a distinct step on the lower mass end. As a result, our choice of Gaussian prior help removes nonphysical HOD without altering any of our main conclusions. 

Finally, we expect the difference from choices of priors to be further reduced as we achieve tighter constraints with more precise measurements using larger samples in the future.

% \clearpage

\section{Author Affiliations}
\label{sec:affiliations}
% List of institutions
$^{1}$Kavli Institute for Particle Astrophysics and Cosmology, Stanford University, 452 Lomita Mall, Stanford, CA 94305, USA\\
$^{2}$SLAC National Accelerator Laboratory, 2575 Sand Hill Road, Menlo Park, CA  94025, USA\\
$^{3}$Department of Physics, Kansas State University, 116 Cardwell Hall, Manhattan, KS 66506, USA\\
$^{4}$Department of Astronomy, The Ohio State University, 140 W. 18th Ave., Columbus, OH 43210, USA\\
$^{5}$Institute of Cosmology $\&$ Gravitation, University of Portsmouth, Dennis Sciama Building, Portsmouth, PO1 3FX, UK\\
$^{6}$Miller Institute for Basic Research in Science, University of California, Berkeley, CA 94720, USA\\
$^{7}$Physics Division, Lawrence Berkeley National Laboratory, Berkeley, CA 94720, USA\\
$^{8}$Department of Physics and Astronomy, University of Utah, 115 South 1400 East, Salt Lake City, UT 84112, USA\\
$^{9}$Institue for Astronomy, Royal Observatory, University of Edinburgh, UK\\
$^{10}$Tata Institue of Fundamental Research, Homi Bhabha Road, Mumbai 400005, India\\
$^{11}$Departamento de Física Teorica, Facultad de Ciencias, Universidad Autonoma de Madrid, ES-28049 Madrid, Spain\\
$^{12}$Centro de Investigacion Avanzada en Física Fundamental (CIAFF), Facultad de Ciencias, Universidad Autonoma de Madrid, ES-28049 Madrid, Spain\\
$^{13}$Lawrence Berkeley National Laboratory, 1 Cyclotron Road, Berkeley, CA 94720, USA\\
$^{14}$Physics Dept., Boston University, 590 Commonwealth Avenue, Boston, MA 02215, USA\\
$^{15}$Dipartimento di Fisica ``Aldo Pontremoli'', Universit\`a degli Studi di Milano, Via Celoria 16, I-20133 Milano, Italy
$^{16}$Department of Physics \& Astronomy, University College London, Gower Street, London, WC1E 6BT, UK\\
$^{17}$Instituto de F\'{i}sica, Universidad Nacional Aut\'{o}noma de M\'{e}xico, Cd. de M\'{e}xico C.P. 04510, M\'{e}xico\\
$^{18}$Departamento de F\'isica, Universidad de los Andes, Cra. 1 No. 18A-10, Edificio Ip, CP 111711, Bogot\'a, Colombia\\
$^{19}$Department of Physics, The Ohio State University, 191 West Woodruff Avenue, Columbus, OH 43210, USA\\
$^{20}$Center for Cosmology and AstroParticle Physics, The Ohio State University, 191 West Woodruff Avenue, Columbus, OH 43210, USA\\
$^{21}$Department of Physics, The University of Texas at Dallas, Richardson, TX 75080, USA\\
$^{22}$Department of Physics, Southern Methodist University, 3215 Daniel Avenue, Dallas, TX 75275, USA\\
$^{23}$Institut de F\'{i}sica d’Altes Energies (IFAE), The Barcelona Institute of Science and Technology, Campus UAB, 08193 Bellaterra Barcelona, Spain\\
$^{24}$NSF's NOIRLab, 950 N. Cherry Ave., Tucson, AZ 85719, USA\\
$^{25}$Instituci\'{o} Catalana de Recerca i Estudis Avan\c{c}ats, Passeig de Llu\'{\i}s Companys, 23, 08010 Barcelona, Spain\\
$^{26}$Department of Physics and Astronomy, Siena College, 515 Loudon Road, Loudonville, NY 12211, USA\\
$^{27}$Department of Physics \& Astronomy and Pittsburgh Particle Physics, Astrophysics, and Cosmology Center (PITT PACC), \\  \hspace{4pt} University of Pittsburgh, 3941 O'Hara Street, Pittsburgh, PA 15260, USA\\
$^{28}$National Astronomical Observatories, Chinese Academy of Sciences, A20 Datun Rd., Chaoyang District, Beijing, 100012, P.R. China\\
$^{29}$Waterloo Centre for Astrophysics, University of Waterloo, 200 University Ave W, Waterloo, ON N2L 3G1, Canada\\
$^{30}$Perimeter Institute for Theoretical Physics, 31 Caroline St. North, Waterloo, ON N2L 2Y5, Canada\\
$^{31}$Space Sciences Laboratory, University of California, Berkeley, 7 Gauss Way, Berkeley, CA 94720, USA\\
$^{32}$Universit\'e Paris-Saclay, CEA, Institut de recherche sur les lois Fondamentales de l'Univers, 91191, Gif-sur-Yvette, France.
$^{33}$Department of Physics and Astronomy, Sejong University, Seoul, 143-747, Korea\\
$^{34}$Centro de Investigaciones Energ\'e ticas, Medioambientales y Tecnol\'o gicas (CIEMAT), Madrid, Spain
$^{35}$Department of Physics, University of Michigan, Ann Arbor, MI 48109, USA\\
$^{36}$Department of Physics and Astronomy, Ohio University, Athens, OH 45701, USA\\
$^{37}$Laboratory of Astrophysics, \'Ecole Polytechnique F\'ed\'erale de Lausanne (EPFL), Observatoire de Sauverny, CH-1290 Versoix, Switzerland\\%%%%%%%%%%%%%%%%%%%%%%%%%%%%%%%%%%%%%%%%%%%%%%%%%%
% Instituto de C`ıencias del Cosmoc, (ICCUB) Universidad de Barcelona (IEEC-UB), Mart´ı i Franqu`es 1, E08028 Barcelona, Spain

% Don't change these lines
\bsp	% typesetting comment
\label{lastpage}
\end{document}